\documentclass[12pt]{article}
\usepackage[latin1]{inputenc}
\usepackage{amsmath}
\usepackage{amsfonts}
\usepackage{amssymb}
\usepackage{graphicx}
\usepackage{geometry}
\usepackage{hyperref}
\usepackage{multirow}
\usepackage[english]{babel}
\usepackage[bottom]{footmisc}
\usepackage{graphicx,algorithmic}
\usepackage[linesnumbered,ruled,vlined]{algorithm2e}
\SetCommentSty{mycommfont}

\title{Multiscale major factor selections for complex system data with structural dependency and heterogeneity}

\author{Hsieh Fushing $^{1}$, Elizabeth Chou $^{2}$, and Ting-Li Chen $^{3,}$*}
\date{}

\begin{document}
\maketitle
\noindent 1 Department of Statistics, University of California, Davis, CA 95616, USA\\
2 Department of Statistics, National Chengchi University, Taipei 11605, Taiwan\\
3 Institute of Statistical Science, Academia Sinica, Taipei 11529, Taiwan\\
*email: tlchen@stat.sinica.edu.tw

%




\begin{abstract}
Based on structured data derived from large complex systems, we computationally further develop and refine a major factor selection protocol by accommodating structural dependency and heterogeneity among many features to unravel data's information content. Information Theoretical measurements are employed in this protocol along global-to-locality paths across different real and simulated complex systems. Two operational concepts: ``de-associating'' and its counterpart ``shadowing'' that play key roles in our protocol, are reasoned, explained, and carried out via contingency table platforms. This protocol via ``de-associating'' capability would manifest data's information content by identifying which covariate feature-sets do or don't provide information beyond first identified major factors to join the collection of major factors as secondary members. Our computational developments begin with globally characterizing a complex system by structural dependency between multiple response (Re) features and many covariate (Co) features. At the global phrase, major factors are selected to characterize the global Re-Co relation. Then the entire data set is divided into structural heterogeneity based localities. At the local phase, various locality-specific Re-Co relations are designed and explored to characterize locality-specific details. We first apply our major factor selection protocol on a Behavioral Risk Factor Surveillance System (BRFSS) data set to demonstrate discoveries of localities where heart-diseased patients become either majorities or further reduced minorities that sharply contrast data's imbalance nature. We then study a Major League Baseball (MLB) data set consisting of 12 pitchers across 3 seasons, and reveal detailed multiscale information content regarding pitching dynamics, and provide nearly perfect resolutions to the Multiclass Classification (MCC) problem and the difficult task of detecting idiosyncratic changes of any individual pitcher across multiple seasons. We conclude by postulating an intuitive conjecture that large complex systems related inferential topics can only be efficiently resolved through discoveries of data's multiscale information content reflecting system's authentic structural dependency and heterogeneity.
\end{abstract}










\section{Introduction}

Studying a large complex system \cite{gellmann,adami,tumer} has become much more feasible in this Big Data era ever than before. This feasibility is primarily provided by the wide-spreading availability of structured databases. Such databases allow data analysts to study wide ranges of complex systems and even discover interesting patterns without domain knowledge or expertise. This phenomenon is somehow unthinkable to domain scientists of related complex systems.

There are at least three attributing factors to this phenomenon. First, technological advances have driven the costs of collecting unstructured data to a very affordable level in terms of time and money, see such as videos and images in YouTube,. Secondly, each structured data set is primarily algorithmically, automatically, or even manually transformed from unstructured data format by experts who have invested efforts of annotation. For instance, each baseball pitch delivered by a Major League Baseball (MLB) pitcher in anyone of MLB's 30 stadiums is recorded by two high-speed cameras, which are algorithmically and automatically annotated into 22 features' measurements coupled with all other related information regarding its batter and batting result. Once this baseball is hit, its trajectory is traced by radars and then categorized with respect to the baseball field. More than half million of such pitches' structured data points are annually stored and made available in the PITCHfx and Statcast database \cite{FC21,FCC21}. Together with the radar data, ``Statcast Arrives, Offering Way to Quantify Nearly Every Move in Game'' (Sandomir, Richard, April 21, 2015, The New York Times.) These two databases can be found in the MLB website. Thirdly, the wide-spreading coverage of Internet that has made structured databases available to everyone with Internet access. For instance, multiple annual survey data sets of more than 400K of people in Behavioral Risk Factor Surveillance System (BRFSS) are available in its website \cite{remington,mokdad,pierannunzi}. Here, MLB and BRFSS are just two known examples among many currently available in Internet.

What is complex system dynamics contained in a structured data set? Here we consider a generic structural formation for a concrete notion about the answer to this qustion. Within such a data set, a data point is in a form of $L+K$D vector of measurements or categories. The first $L$ components are derived from the response (Re) features denoted as ${\cal Y}=(Y_1, ..., Y_L)'$, and the rest of $K$ components are derived from $K$ one-dimensional covariate (Co) features denoted as $\{V_1,...,V_K\}$. Some or even all response and covariate features could be categorical. It is natural to imagine the collective of associative relationships between response features and covariate features as the Re-Co dynamics. This dynamics of interest is then taken as the chief characteristic of the complex system under study.

Only for expositional concreteness, we further give an implicit functional expression of such Re-Co dynamics via a collection of $M$ unknown constituent mechanisms, $\{{\cal F}_m\{A^*_m\}|m=1,.., M\}$, that are governed by a also unknown global function ${\cal G}(\cdot)$ in the following fashion:
  \begin{align}
  {\cal Y} &=\begin{bmatrix}
           Y_{1} \\
           Y_{2} \\
           \vdots \\
           Y_{L}
         \end{bmatrix}\cong {\cal G}({\cal F}_1\{A^*_1\},{\cal F}_2\{A^*_2\},..{\cal F}_M\{A^*_M\}, \oplus \varepsilon).
  \end{align}
This layout is motivated by the fact that a complex system naturally has mutliscale deterministic and stochastic structures \cite{anderson,crutchfield}. Here, $M$ is unknown. And we neither have any priori knowledge, nor assumptions about the $M$ functional forms of ${\cal F}_m\{\cdot\}$. Each ${\cal F}_m\{\cdot\}$ is a mechanism component of the whole dynamics governed by an unknown and unspecified structural function ${\cal G}(\cdot)$. The $\oplus \varepsilon$ within ${\cal G}(\cdot)$ simply means that various different forms of unknown stochasticity exist across all $\{{\cal F}_m\{A^*_m\}|m=1,.., M\}$ upon unspecified multiple scales embedded within ${\cal G}(\cdot)$. This fact has been well-recognized in Nobel physicist P. W. Anderson's 1972 Science paper with title:``More is different'' \cite{anderson}.

Basically, we only know that the data curator has chosen one response feature set ${\cal Y}$  against one covariate feature set $\{V_1,...,V_K\}$ from a complex system of interest. Our ultimate goal here is to explicitly identify all involving covariate feature subset $\{A^*_m|m=1,.., M\}$ without knowing any specifications of local mechanisms $\{{\cal F}_m\{A^*_m\}|m=1,.., M\}$, nor the global governing structure ${\cal G}(\cdot)$. We term such a covariate feature subset $A^*_m$ as a major factor of the Re-Co dynamics. As would become clear in the latter part of this paper, the knowledge of this collection of major factors $\{A^*_m|m=1,.., M\}$ indeed will be sufficient enough to weave the data's multiscale information content, which will sustain almost all inferential resolutions and decision-makings, like predictive and testing ones. To a great extent, this seemingly is a surprising statement at this point, but it will become an intuitive and obvious one later.

It is worth mentioning that the $L+K$ features selected by its data curators likely give rise to unexpected exquisite information content that is not previously known even to the data curators themselves regarding a targeted real-world complex system. That is, data analysts can discover knowledge from a structured database about a targeted complex system that goes far beyond the curator's intelligence. Since the associative relations among all involved features: response (Re) and covariate (Co), are usually rather too complex to be fully grasped by individual scientists because heterogeneity, nonlinearity, and unknown structural dependency are involved across multiple scales. That is, when these features coherently exhibit multiscale pattern information, such multiscale heterogeneous and nonlinear pattern information is unlikely fully known.

In this paper, our goal here is to further develop Categorical Exploratory Data Analysis (CEDA) aided by an expanded major factor selection protocol to become a computing paradigm that has potentials to be adaptable to diverse complex system's specifications for unraveling its data's information content. This revised computational paradigm is developed and illustrated through simulated examples, and then applied onto baseball pitching and human heart disease dynamics from Statcast of MLB and a Kaggle version of BRFSS structured databases, respectively. We demonstrate that the end products of such applications indeed offer comprehensive data's information content of multiscale, heterogeneous and nonlinear nature. We further show that such computed information content indeed provides resolutions to multiple topic issues in statistics and machine learning as byproducts.

\subsection{Brief reviews of Information Theoretical measurements and previous works.}
In contrasting to the focus of multiscale structural dependency and heterogeneity here, our previous major factor selection protocol operates on a single layer fashion and mainly works for more or less stochastically independent covariate features \cite{CCF22a,CCF22b}. Our focal issue here stems from the observed fact that, when several important covariate features are highly associated, their individual and joint interacting effects via conditional mutual information are vastly convoluted and intertwined. If these highly associated features were taken as being independent, then their true effects are likely either ignored or mistaken. We illustrate this motivating key point from the Theoretical Information perspective.

Let the categorized or categorical response variable be denoted as ${\cal Y}$ that originally could involve multiple features of any data types: continuous, discrete or categorical or their mixed. It is noted that a continuous feature can be coherently categorized with respect to its own histograms \cite{hsiehroy}, and multiple categorical or categorized features could always be fused into one single variable via a collection of occupied multidimensional hypercubes. Likewise for any covariate feature-set of any sizes of any data types.  Henceforth, we make use of capital letters $A$ or $B$ to denote different subsets of categorized or categorical covariate features, and to simultaneously denote different categorical variables derived by fusing correspondingly different subset of categorical or categorized covariate features.

Then, we explicitly construct a contingency table for each pair of categorical variables like $({\cal Y}, A)$, $({\cal Y}, B)$,  $(A, B)$ and $({\cal Y}, (A, B))$. For instance, we denote the contingency table for $({\cal Y}, A)$ by $C[A-vs-{\cal Y}]$ with categories of $A$ and ${\cal Y}$ being arranged along the row- and column-axes, respectively. Likewise for all other pairs. As a convention in this paper, we typically arrange categories of ${\cal Y}$ on column-axis. Along the row-axis, each row of $C[{\cal Y}-vs-A]$, says $A=a$, defines a conditional multinomial random variable with a conditional (Shannon) entropy (CE) denoted by $H[{\cal Y}|A=a]$. We then calculate the expected CE $H[{\cal Y}|A]$ as a properly weighted sum of the collection of row-wise CEs $\{H[{\cal Y}|A=a]\}$. The marginal column-wise and row-wise entropies are denoted as $H[{\cal Y}]$ and $H[A]$, respectively.

It is known that $H[{\cal Y}|A]$ conveys the expected amount of remaining uncertainty in ${\cal Y}$ after knowing $A$. In reverse, by knowing ${\cal Y}$, $H[A|{\cal Y}]$ conveys the expected amount of remaining uncertainty in $A$ after seeing ${\cal Y}$. The two conditional entropy drops, i.e. differences $H[Y]-H[{\cal Y}|A]$ and $H[A]-H[A|{\cal Y}]$, indicate the shared amount information between $A$ and ${\cal Y}$:
\begin{eqnarray*}
H[{\cal Y}]-H[{\cal Y}|A]&=&H[A]-H[A|{\cal Y}]\\
&=&H[A]+H[Y]-H[A,{\cal Y}]\\
&=&I[{\cal Y};A].
\end{eqnarray*}
where $I[{\cal Y};A]$ denotes the mutual information between ${\cal Y}$ and $A$.

Next, we consider the mutual information between the bivariate $(A,B)$ and ${\cal Y}$ starting from their conditional mutual information as:
\[
I[A; B|{\cal Y}]=H[A|{\cal Y}] +H[B|{\cal Y}]-H[(A,B)|{\cal Y}].
\]
Further, we decompose this mutual information into the following two key components: 1) the sum of individual CE-drops of $A$ and $B$ and 2) the difference of conditional and marginal mutual information of $A$ and $B$:
\begin{eqnarray*}
H[{\cal Y}]-H[{\cal Y}|(A, B)]&=&H[(A, B)]-H[(A,B)|{\cal Y}];\\
&=&H[A]+H[B]-I[A;B] - \{H[A|{\cal Y}] + H[B|{\cal Y}]-I[A;B|{\cal Y}]\};\\
&=&\{H[{\cal Y}]-H[{\cal Y}|A] + H[{\cal Y}]-H[{\cal Y}|B]\}+\{I[A;B|{\cal Y}]-I[A;B]\}.
\end{eqnarray*}

The above decomposition precisely conveys the essence of interpretable meaning of conditional mutual information when the two involving feature sets $A$ and $B$ are indeed marginally independent because $I[A;B]=0$. And if $I[A;B|{\cal Y}]$ is relatively large, then we are certain that $A$ and $B$ have significant interacting effect in reducing the uncertainty of ${\cal Y}$. However, if $A$ and $B$ are indeed highly associated or dependent, then $I[A;B] >0$, then the last term of the above equation: $\{I[A;B|{\cal Y}]-I[A;B]\}$, can be negative. We then face two chief difficulties: 1) it is hard to determine whether the smaller CE-drop by including either $A$ or $B$ is significant or not; 2) it is hard to assess whether the $A$ and $B$ have significant interacting effect or not even when $\{I[A;B|{\cal Y}]-I[A;B]\}$ is positive. We make simulated examples to explicitly demonstrate such difficulties in the next section below.

\subsection{Present computational difficulties.}
Due to such difficulties, Theoretical Information Measurements based criterion of major factor selection developed in \cite{CCF22a} likely dismisses the important contributions from one of the two features together with their joint interacting effect. Such a computed dismissal decision is apparently neither realistic, nor valid in dealing with wide ranges of real-world complex systems.

That is, we definitely need better understandings on teasing out complicate effects among associated covariate features, and simultaneously need precise improvements on computational methodologies in order to differentiate and evaluate what are effects of features or feature-sets beyond the effect of any focal feature or feature-set. Such conceptual as well as computational developments make up the technical theme topic in this paper.

One new concept of ``shadowing'' and one old concept of ``de-associating'' are devised between two categorical features or feature-sets, say $A$ and $B$. These concepts are implemented on a platform of contingency table. Upon $C[A-vs-B]$, we construct a new categorical variable $B^*[A]$, called $B$ shadowed by $A$, equipped with the following properties: 1) $B^*[A]$ and $B$ have the same marginal distribution; 2) $A$ and $B^*[A]$ retains the same association as $A$ and $B$; 3) but $B^*[A]$ is a composition of the projected part of $A$ on $B$ together with a independent replicated part that is distributed same as the part of $B$ being independent of $A$. The indirect utility of $B^*[A]$ is that it allow us to see what information of $A$ is in $B$?

Contrasting with shadowing, the de-associating is the operation that explicitly extract the part of $B$ being independent of $A$. This operation and its function of de-associating is just conditioning, which is a well-defined mathematical concept, but very hard to see explicitly. Nevertheless, the conditional variable $B$ given $A$ has a very simple structure when $A$ is indeed categorical. That is, under the platform $C[A-vs-B]$, this conditional random variable is explicitly specified by the row-wise Multinomial randomness, collectively. For this concrete appearance under categorical setting, this de-associating operation is specifically denoted $B^{\bot}[A]$.

Further, if all features, including the possibly multiple dimensional response ${\cal Y}$, are commonly made to be de-associating with respect to $A$, then the entire data set is divided with respect to each category of $A$, respectively. Upon each such data subset, in which $A$ is fixed at a constant category, we can evaluate and reveal information provided by all other features, which become less or much less associativity dependent among each other within localities defined by categories of $A$. All these pieces of information are independent of (beyond) what $A$ can provide. This recognition and its functionality of $B^{\bot}[A]$ upon the contingency table platform are especially important and useful for us to access individual as well as interacting effects among highly associated features. We explicitly demonstrate the functionality of  $B^{\bot}[A]$ in both real complex systems of pitching dynamics and heart disease.

In summary, our computational developments in this paper not only motivate critical computing steps for building a major factor selection protocol, but also explicitly show how to extract essential as well as intricate pattern information regarding which features play which roles under the setting of having covariate features being heavily dependent. This new protocol together with original protocol proposed  in \cite{CCF22a,CCF22b}, which is valid for settings with less assertively dependent covariate features, such as within localities defined by major factors, will constitute a comprehensive paradigm for analyzing Re-Co dynamics involving with structural dependency among all covariate features and structural heterogeneity in information content within any complex system study.

This paper is organized as follows. The Introduction section is followed by the Section 2 where we develop and illustrate our major factor selection protocol via simulated examples. The Sections 3 and 4 are devoted to analyzing two complex systems of heart disease and MLB pitching dynamics, respectively. The Conclusion section contains several remarks, implications and future research directions and problems.

\section{Computational developments for major factor selection along with motivating examples under highly associative settings}
In this section, we develop our major factor selection protocol based on ``shadowing'' and ``de-associating'' operations upon contingency table platform. We begin by considering two linearity based model-examples to explicitly illustrate these two operations, respectively. Through the first example, we demonstrate that ``Shadowing'' provides analytic views of information contributed by any major factors or feature-sets. The second example is slightly modified version of the first one with the specific aims of demonstrating how ``de-associating'' can computationally confirm whether any major factor candidates could or couldn't provide information beyond the confirmed major factors.

The organization of this computational development section is given as follows. In first subsection, we lay out the two model-examples, upon which we also report some instability related computational phenomena resulted from LASSO. These settings and results here are meant to be contrasting with, but not intended to contest the consistency results of LASSO in statistics and machine learning literatures \cite{meier,meinshausen,zhao}. In the 2nd subsection, under the first example, we report CE and CE-drop (conditional mutual information) and identify potential major factors, and then ``shadowing'' and ``de-associating'' operations are carried out in order for confirming a collection of major factors. In the 3rd subsections, under the second example, we also report CE and CE-drop (conditional mutual information) and identify potential a candidate collection of major factors and then ``de-associating'' operation is successively carried out to confirm that there exist no other major factors being able to offer information beyond this collection of major factors. In the last subsection, we explicitly lay out our major factor selection protocol with operational remarks.

\subsection{Two illustrating examples with highly associated covariate features}
Our first illustrative example is specified by the following linear Re-Co dynamics constituted by 7 highly associated 1D features (or variables) given as follows:
\begin{eqnarray*}
Y&=&0.8 X_1 + X_2 + 1.2 X_3 + X_{11} +\epsilon,\\
X_7&=& (X_1+X_2+X_3+X_4+X_{11})/3.66,\\
X_i &\sim& N(0,1), \; \rho(X_i, X_j)=0.7, \; i,j=1,..,6;\\
X_k &\sim& N(0,1), i.i.d,\; k=8,...,11,\\
X_{11}&:& \textrm{is the only unobserved hidden variable.}
\end{eqnarray*}
Due to the presence of unobserved hidden variable $X_{11}$, it is reasonable to say at least intuitively that $\{X_7, X_4\}$ is the primary order-2 major factor of the Re-Co dynamics. Since $Y=X_7-X_4+ 0.2(X_3-X_1)+\epsilon$. It is also intuitive that extra and delicate evidences are certainly needed in order to expand $\{X_7, X_4\}$  into a collection of major factors  $\{\{X_7, X_4\},X_3, X_1\}$. In this subsection, we perform the classic LASSO approach under this linear Re-Co dynamics to prepare for comparisons with what our major factor selection can do in the next two subsections.

\begin{figure}
 \centering
 \includegraphics[width=6in]{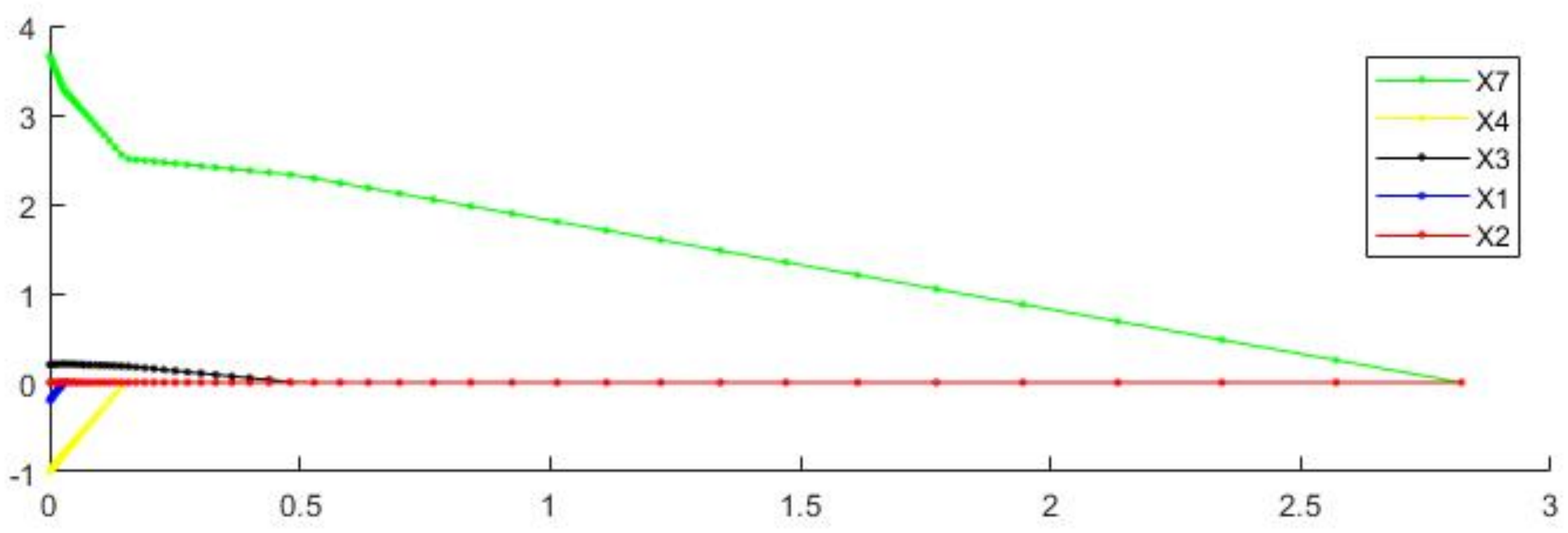}
 \caption{Results of LASSO estimations with respect to L-1 penalty $\lambda$ on X-axis.}
 \label{exp20220304}
 \end{figure}

Based on a simulated data set with a sample size of 100K, LASSO approach is conducted and results of parameter estimations with respect to L-1 penalty $\lambda$ are reported in Figure~\ref{exp20220304}. It is evident that the least square estimation (with $\lambda=0$) is able to provide the exact and precise model structure. But, as $\lambda$ increasing from zero, though the presence of $X_7$ is persistent, its estimated values are shrinking. Overall LASSO results show diminishing importance of $X_4$, $X_3$ and $X_1$. In summary, even when $\lambda$ is only slightly positive, all resultant structures are rather off the true structure.

It is worth emphasizing that the above results rely heavily on the priori knowledge of linear structure. Nonetheless, severely biased results are still concluded. Therefore, it is crucial to know what can be discovered when no such knowledge being available? As would be developed and seen in the next two subsections, comprehensive resolutions for this essential question and subsequently raised questions below would merge.

\begin{figure}
 \centering
  \includegraphics[width=5in]{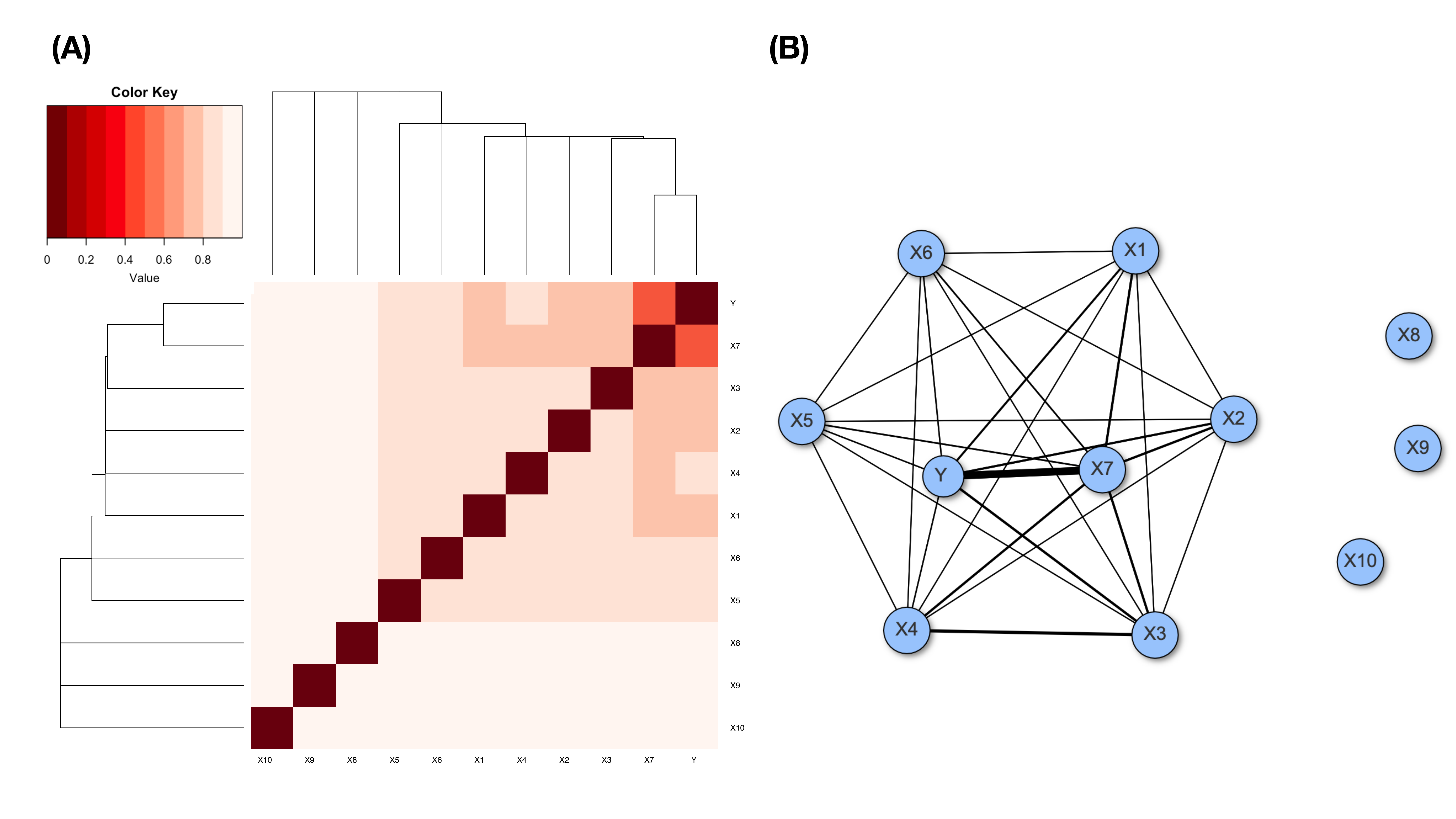}
 \caption{MCE heatmap and network of 12 features involving in the illustrative example.}
 \label{fig:20220304MCE}
 \end{figure}

It is typical that the first batch of pieces of information in a structured data set are revealed through its associative heatmap, graph or network among these 12 variables as seen in Figure~\ref{fig:20220304MCE}. This heatmap and its corresponding network apparently contain a complete clique among variables in $\{Y, X_1,..., X_7\}$ due to their strong pairwise correlations, and a triplet of isolated nodes of $X_8$, $X_9$ and $X_{10}$. Such pieces of information, such as communities \cite{chenhsieh}, are beneficial to know. The complete clique indicates that the Re-Co dynamics likely involves with hard to untangled relations. Though $X_7$ evidently plays a dominant role in the Re-Co dynamics, while $X_4$ plays a ``negative'' role within $X_7$ and not directly involved in $Y$. Can we detect this fact?

Further, while $X_1$, $X_2$ and $X_{3}$ have slightly increasing coefficients in the linear structure, their roles are not increasing important. Since $X_2$ is ``covered'' by $X_7$ in the sense that $X_2$ is not important at the presence of $X_7$. Further, though $X_1$ and $X_3$ are ``equal'' in the linear structure, their roles of reducing uncertainty of $Y$ might not be equal. Furthermore, these three features are highly associated. Can we differentiate their intricate differences? On the other hand, $X_5$ and $X_6$ play no roles in the Re-Co dynamics, but they are highly correlated (under normality) with $\{X_1, X_2, X_3, X_4, X_7\}$. Can we tease out their roles from the roles played by the 5 directly and 5 indirectly involving covariate features.

Finally, it is clear that the isolated feature or variable nodes $\{X_8, X_9, X_{10}\}$ likely do not play any relational roles in the Re-Co Dynamics. But in real-world data, we need to take into account the potentials that such features could join other features to form essential high-order interacting roles. Therefore, we need to make sure whether such interacting relations exist in data or not. Though they play no roles in the Re-Co dynamics, these three features indeed computationally serve as baselines for our Shannon entropy evaluations when attempt to keep effects of the finite sample phenomenon or so-called curse of dimensionality at bay, see \cite{FCC23} for practical guidelines on CE and mutual information evaluations.

Next our second illustrative example is a slightly modified version of the first example. Though, this simpler version would reiterate the cause of instability of LASSO within settings that contain two or more ``wells'' of locally optimal solutions in the optimizing landscape, it is primarily designed to showcase that no information regarding $Y$ could be found beyond a chief collection of major factors.

\begin{figure}
 \centering
 \includegraphics[width=6in]{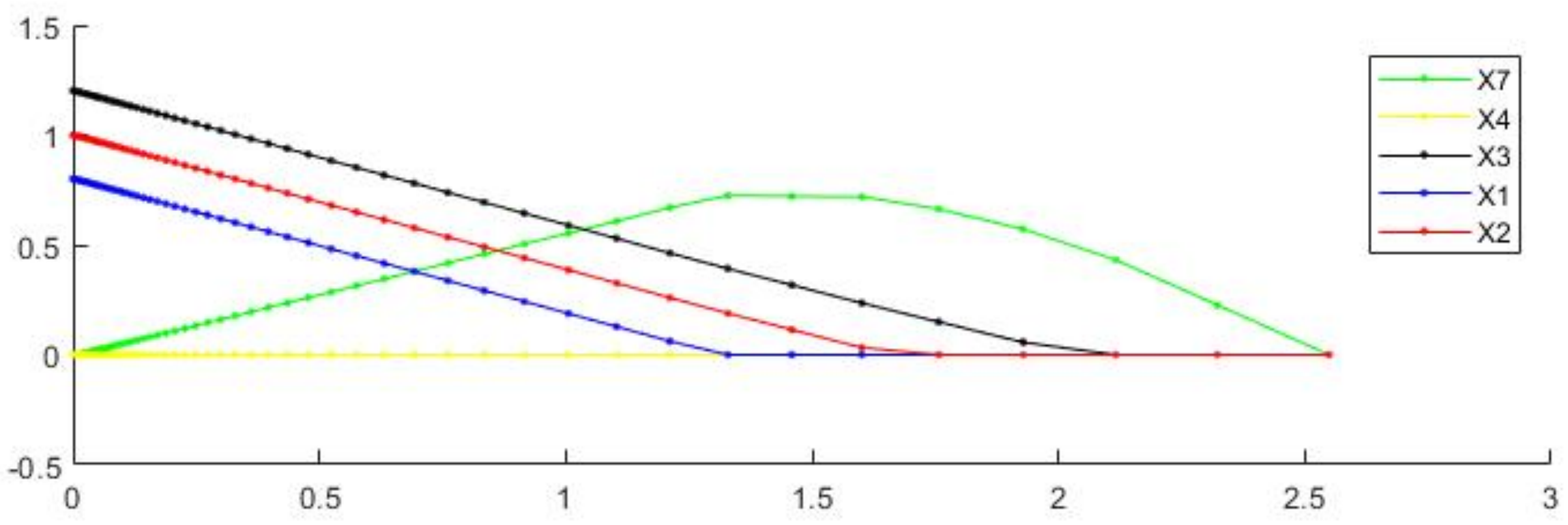}
 \caption{Results of LASSO estimations of modified example with respect to L-1 penalty $\lambda$ on X-axis.}
 \label{exp20220408}
 \end{figure}

In this example, we take off the hidden factor $X_{11}$ in $Y$ as follow:
\begin{eqnarray*}
Y&=&0.8 X_1 + X_2 + 1.2 X_3+\epsilon,\\
X_7&=& (X_1+X_2+X_3+X_4+X_{11})/3.66,\\
X_i &\sim& N(0,1), \; \rho(X_i, X_j)=0.7, \; i,j=1,..,6;\\
X_k &\sim& N(0,1), i.i.d,\; k=8,...,11,\\
X_{11}&:& \textrm{is an unobserved hidden variable}.
\end{eqnarray*}
In this second example, the feature-pair $\{X_4, X_7\}$ apparently becomes a less effective order-2 major factor candidate than the collection of three order-1 major factors: $\{X_1, X_2, X_3\}$. This essential fact would be reported based on CEs computations in the 4th and 5th subsections below. As a by-product, again the LASSO results with respect to a range of penalty $\lambda$ are depicting instable and biased results as presented in Figure~\ref{exp20220408}.

The Figure~\ref{exp20220408}delivers an even clearer message than that delivered by Figure~\ref{exp20220304}:``the penalty is not intrinsic in these linear regression settings". The collection of major factors: $\{X_1, X_2, X_3\}$, is clearly needed to be held constant at least for a small range of $\lambda$ for all practical and realistic reasons. But this pattern does not happen. The implication of such results is simply: LASSO is hardly a realistic tool for extracting information even under the man-made linearity when its optimizing landscape contains multiple local minimum, which is realistically likely to be the case in majority of real-world complex systems.

\subsection{Major factor selection protocol for Example-1.}
In our previous works \cite{CCF22a,CCF22b}, our major factor selection protocol was proposed to discover and identify varying orders of major factors based on two criterions: [C1:confirmable] and [C2:irreplabeable], which are explicitly listed in the last subsection of this section. These two criterions operate under the assertion that CE evaluations or estimations are persistently stable throughout all vital feature-settings. In reality, CEs would vary only slightly downward as ``dimensions'' of contingency tables expanding. But, when contingency tables' dimensions grow too big with respect to the sample size, then the effects of so-called finite sample phenomenon or curse of dimensionality would kick in, that is, all feature-sets work like random noise variables. To guard against such finite sample phenomenon, one key characteristic of our major factor selection protocol is that, according to [C1:confirmable], a candidate of major factor must achieve a significant larger CE-drop, which a mutual information estimate of this major factor candidate and the response variable ${\cal Y}$, than that of a random noise variable under the same dimension of contingency table. This requirement is striking distinct to the one requiring consistent CE estimate. The rigorous discussion of related issues of how to make reliable CE-drop evaluations is given in \cite{FCC23}, where results are sharply contrasting with results reported in \cite{paninski}.

As such, a potential major factor candidate's CE-drop must significantly larger than a random noise variable's under the same dimension of contingency table, which is termed the criterion [C1:confirmable] proposed and used in the selection protocol developed in \cite{CCF22a,CCF22b}. We show such implementations of criterion [C1:confirmable] in Table ~\ref{table20220304} that report calculated Conditional Entropy(CE) of feature-sets ($CE[Y|\{X_i,..,X_j\}]$) across five settings: 1-feature to 5-feature settings. Across these five settings, the dimensions of contingency tables grow with the number of members in $\{X_i,..,X_j\}$. Even though, theoretically $CE[Y]=CE[Y|\{X_i,..,X_j\}]$ when $Y$ is independent of $\{X_i,..,X_j\}$ disregarding the number of members in $\{X_i,..,X_j\}$, the empirical fact, as shown in Table ~\ref{table20220304}, is that the numerical values of $CE[Y|\{X_i,..,X_j\}]$ would slightly decrease as the number of members of $\{X_i,..,X_j\}$ increasing. Nonetheless, the decreasing amounts is relatively small if average cell count in the corresponding contingency table remains 10 or more \cite{FCC23}. Thus, we compare CEs within the same feature-setting, not cross over different feature-settings.

With the above notes in mind, we summarize computed patterns in Table ~\ref{table20220304} as follows. Starting from the 1-feature setting, one estimated entropy of $Y$ is accordingly calculated as $CE[Y|X_{10}]=2.4091$ to reflecting the dimensionality of  contingency table and stochastic independence of $Y$ and $X_{10}$. The feature $X_7$ achieves the lowest CE: $1.0498$, so its CE-drop is calculated as $1.3593=2.4091-1.0498$, which is an estimate of mutual information $\hat{I}^{(1)}[Y, X_7]$. The superscript indicates that the estimated mutual conditional information is calculated within the 1-feature setting. $X_3$, $X_2$ and $X_1$ achieve the top 2nd, 3rd and 4th ranked CE-drops that are significantly larger than the CE-drop of $X_4$, which is calculated as $\hat{I}^{(1)}[Y, X_4]=2.4091-2.0478=0.3613$.

On the 2-feature setting, one estimated entropy of $Y$ is calculated as $CE[Y| X_8, X_{9}]=2.4014$. It is known that $X_7-X_4$ is very close to the functional structure of $Y$. Since both embrace the almost same linear structure and the common hidden factor $X_{11}$. Thus, it is as expected that the feature-pair $\{X_4,X_7\}$ achieves the lowest CE with an estimated mutual information (CE-drop):
\[
\hat{I}^{(2)}[Y, \{X_4,X_7\}]=2.4014-0.7648=1.6366.
\]
Nonetheless, this CE-drop of $\{X_4,X_7\}$ is smaller than the sum of individual CE-drops of $X_7$ and $X_4$, which is equal $1.7206$. This fact indicates that the conditional mutual information of $\{X_4,X_7\}$ given $Y$ is less than the marginal mutual information of $\{X_4,X_7\}$, that is,
\[
\hat{I}^{(2)}[\{X_4,X_7\}|Y] -\hat{I}^{(2)}[\{X_4,X_7\}]=-0.0840 < 0.
\]
In other words, the so-called ecological effect, which plays the key role in the 2nd criterion [C2:irreplabeable] used in \cite{CCF22a,CCF22b} under the independence setting, is not seen in this case of having stochastically highly dependence between $X_7$ and $X_4$. However, $\{(X_4,X_7)\}$ is an evident order-2 major factor in the Re-Co dynamics of $Y$ in this illustrative example. Further, the three feature-pairs $\{X_1,X_2\}$,  $\{X_1,X_3\}$ and $\{X_2,X_3\}$, like $\{X_4,X_7\}$, do not have ecological effects even though they are parts of linear structures of $Y$. Furthermore, related issues pertaining to the two feature-pairs: $\{X_1,X_7\}$ and $\{X_3,X_7\}$, are somehow more intricate. Since we see that they almost bear no improvements upon $X_7$, while they are supposed to reveal some intricate degrees of improvements according to the linear structure of $Y$. Also, from the 3-feature setting, the two triplets:$\{X_1,X_4, X_7\}$ and $\{X_3,X_4, X_7\}$, show very limited improvements upon $\{X_4,X_7\}$. That is, the subtle, but visible effects of adding $X_1$ or $X_3$ to expand the collection of major factor $\{(X_4,X_7)\}$ are not seen.

In summary, clearly all above observed issues rest heavily on the structural dependency among covariate features. That is, we need better fundamental understanding to advance the categorical exploratory data analysis (CEDA) in achieving the goal: ``Correctly extracting relevant information in data via major factor selection as a way of shedding authentic lights on any complex system study''. As would be developed in the next two sub-subsections, from two unique operational perspectives pertaining to contingency table, we demonstrate that our expanded version of major factor selection protocol could have potentials to be highly adaptable to widely distinct complex systems, and at the same time can shed lights on all aforementioned issues. These two new operations discussed in the next two subsection will enable us to visualize and evaluate subtle effects under the shadow of heavy dependence among covariate features.

\begin{table}[]
\resizebox{\textwidth}{!}{%
\begin{tabular}{llllllll}\hline
1Feature & CE	 &2Feature   &CE	  &3Feature	   & CE	   	& 4Feature	&CE \\ \hline
X7&1.0498&X4\_X7&0.7648&X3\_X4\_X7&0.7048&X3\_X4\_X7\_X10&0.6403\\
X3&1.7927&X3\_X7&1.0152&X1\_X4\_X7&0.7487&X4\_X7\_X8\_X10&0.6524\\
X2&1.8509&X2\_X7&1.0419&X2\_X3\_X7&0.9948&X2\_X3\_X4\_X7&0.6532\\
X1&1.8988&X1\_X7&1.0459&X7\_X8\_X9&1.0187&X4\_X5\_X6\_X7&0.6937\\
X6&2.0473&X2\_X3&1.5503&X2\_X7\_X9&1.0212&X7\_X8\_X9\_X10&0.7934\\
X4&2.0478&X1\_X3&1.5984&X1\_X2\_X3&1.4122&X5\_X7\_X8\_X9&0.8492\\
X5&2.0497&X1\_X2&1.6505&X1\_X3\_X4&1.5459&X1\_X2\_X3\_X7&0.9219\\
X9&2.4089&X3\_X9&1.7864&X3\_X4\_X5&1.6514&X3\_X8\_X9\_X10&1.2531\\
X8&2.4090&X4\_X5&1.9326&X4\_X5\_X6&1.8291&X3\_X4\_X5\_X6&1.4262\\
X10&2.4091&X8\_X9&2.4014&X8\_X9\_X10&2.3072&X4\_X5\_X6\_X8&1.5000\\
\hline
\end{tabular}%
}
\caption{Example-1 with  $N=10^5$. Each categorized 1-features has 12 bins, so a $k$-feature setting has $(12)^k$ $k$D hypercubes.}
\label{table20220304}
\end{table}

\subsubsection{The operation of ``Shadowing''}
In this sub-subsection, we introduce a concept of ``shadowing'' on categorical variables. It allows us to explicitly check and evaluate how much a feature-set $B$ can contribute beyond what feature-set $A$ can offer to a targeted Re-Co dynamics. Its chief function will illuminate the unique characteristics of using contingency table as a computational platform for exploring data's information content.

The shadowing concept is specifically defined and operated on a contingency table built by any pair of categorical feature-sets. Let $A$ and $B$ be the two feature-sets with their contingency table denoted as $C[A-vs-B]$. It is noted that the $C[A-vs-B]$ is built based on a data matrix of ${\cal M}[Y:A:B]$ with its $i$th row being the observed data point $(Y_i, A_i, B_i)$. That is, the data matrix of ${\cal M}[Y:A:B]$ is constructed with $Y$ and $A$ and $B$ being arranged along the column-axis.

Under the Re-Co dynamics of $Y$, a new categorical variable called $B$ shadowed by $A$, denoted as $B^*[A]$, is defined by a newly simulated data matrix of ${\cal M}[Y:A:B*[A]]$. The $i$th row-vector $(Y_i, A_i, B_i)$ of ${\cal M}[Y:A:B]$ is replaced by a new row-vector $(Y_i, A_i, B*[A]_i)$. Here, $B*[A]_i$ is a simulated category of $B$ via the Multinomial distribution $Mn[1, Pro_{A_i}]$ with $Pro_{A_i}$ being the proportional vector of the $A_i$th row of $C[A-vs-B]$. The essence of shadowing is that the category $B*[A]_i$ is simulated with respect to the knowledge of the category $A_i$ of $A$, but without involving with observed $Y_i$. That is, the part of $B$ being ``orthogonal'' to $A$ is taken away row-by-row. It is noted that the contingency table $C[A-vs-B^*[A]]$ constructed based on data matrix of ${\cal M}[Y:A:B*[A]]$ is very much like $C[A-vs-B]$. This computational operation of shadowing is closely related to the mimicking developed in \cite{hsiehchou21}. Consequently, the associative relation of $A$ and $B$ is retained by the associative relation $A$ and $B^*[A]$.

On the other hand, the associative relation between $B^*[A]$ and $Y$ is equal to the part of $B$ being ``parallel'' to $A$ and $Y$. That is, the associative relation of $Y$ and $B$ is weakened because the associative relation of $Y$ with the part of $B$ that is orthogonal to $A$ is gone. As would be described in detail in the next subsection of ``de-associating'', this orthogonal part of $B$ is denoted by $B^{\bot}[A]$, which is Mutlinomial distributed as $Mn[n_{A_i}, Pro_{A_i}]$ with $n_{A_i}$ being the $A_i$th row sum of $C[A-vs-B]$. The capacity of having explicit characterizations of $B^*[A]$ and $B^{\bot}[A]$ is a great advantage pertaining to contingency table platform specifically.

We give an illustrations for this shadowing operation on variables used in the 1st illustrating example in the first subsection. Consider $A=X_7$ and $B=Y$ under the categorized setting. Likewise we generate the data of $Y^*[X_7]$. We use $Y^*[X_7]$ as the response variable with respect to the same collection of covariate features: $\{X_1,..,X_10\}$. The CEs (and CE-drops) are calculated across 1-feature to 3-feature settings in Table~\ref{tableYbyX7}.

\begin{table}[h!]
\centering
\begin{tabular}{llllll}\hline
1Feature & CE	 &2Feature   &CE	  &3Feature	   & CE	   \\ \hline
X7&1.0523&X1\_X7&1.0500&X7\_X8\_X9&1.0207\\
X3&1.8838&X2\_X7&1.0502&X1\_X4\_X7&1.0356\\
X2&1.8846&X4\_X7&1.0503&X2\_X3\_X7&1.0356\\
X1&1.8871&X3\_X7&1.0504&X1\_X2\_X3&1.5115\\
X4&1.8859&X2\_X3&1.6692&X2\_X3\_X4&1.5119\\
X6&2.0339&X1\_X2&1.6708&X1\_X2\_X5&1.6063\\
X5&2.0359&X1\_X3&1.6726&X4\_X5\_X6&1.7052\\
X8&2.4090&X4\_X5&1.7921&X2\_X6\_X8&1.7371\\
X10&2.4090&X3\_X9&1.8775&X3\_X8\_X9&1.8113\\
X9&2.4091&X8\_X9&2.4014&X8\_X9\_X10&2.3066\\
\hline
\end{tabular}%
\caption{Example-1 with  $N=10^5$ and response $Y^*[X_7]$. Each categorized 1-features has 12 bins, so a $k$-feature setting has $(12)^k$ $k$D hypercubes.}
\label{tableYbyX7}
\end{table}

In the 1-feature setting in Table~\ref{tableYbyX7}, the CE of $X_7$ almost retains its CE when the response is $Y$. It is strikingly evident that, with $Y^*[X_7]$ as the response, the CEs of 10 covariate features $X_k$ with $k=1, .., 10$ indeed reflect exactly the linear structure of $X_7$. This phenomenon is indeed for the namesake. Further, we see that no other features or feature-sets can be coupled with $X_7$ to achieve significantly improved CEs in the 2-feature and 3-feature settings. In particular, the feature pair $\{X_4, X_7\}$ achieves a much higher CE (=1.0503) than its CE (=0.7648) when the response variable is $Y$. This fact is strikingly different from the case of having the response variable $Y$ reported in Table ~\ref{table20220304}.  It is also noted that, while CEs of random noise variables $X_8$, $X_9$ and $X_{10}$ remain the same, the feature-triplet $\{X_1, X_2, X_3\}$ achieves a higher CE (=1.5115) than its original CE (=1.4122) with $Y$ as response variable. This CE difference indeed confirm that the triplet $\{X_1, X_2, X_3\}$ can offer extra information beyond $X_7$ on $Y$ because something is lost after the shadowing via $X_7$. By putting these facts together, we conclude that some information about $Y$ has gone missing in $Y^*[X_7]$. In other words, we project that the variable $Y^{\bot}[X_7]$ still share a significant amount of information with all covariate features $X_k$ with $k\neq 7$ as would become more explicit in the next subsection of ``de-associating''.

\subsubsection{The operation of de-associating.}
Here we consider two feature-sets, say $A$ and $B$. We want to explicitly bring out the part of $B$ being ``orthogonal'' to $A$, denoted as $B^{\bot}[A]$. This $B^{\bot}[A]$ is marginally and locally ``de-associated'' with $A$. Here the `` local de-association'' is meant to take all $A$-related associative relational patterns out of $B$ when $A$ is held at a constant $A=a$. Then, with respect to the probabilities $Pr[A=a]$, we define de-associated $B^{\bot}[A]$.

To make such de-associating concept explicit and concrete, we construct $B^{\bot}[A]$ upon the contingency table $C[A-vs-B]$. Given that $A=a$, then the $a$th row of $C[A-vs-B]$ defines a Multinomial random variable $M(1, P^{(B)}_a)$ with probability vector $P^{(B)}_a$ being estimated by the corresponding row-vector of proportion. The randomness of $M(1, P^{(B)}_a)$ is what $B^{\bot}[A]$ is about at $A=a$. That is, data's information of $B^{\bot}[A]$ at $A=a$ is exclusively contained in the observed data subset ${\cal M}[A=a]$. Though, $B^{\bot}[A]$ is simply the conditional random variable of $B$ given $A$, it is worth emphasizing that $B^{\bot}[A]$ can be easily, explicitly and precisely constructed based on the contingency table framework. This recognition is a natural perspective of contingency table, and its construction is one of great merits of contingency table as computational platform.

Here we illustrate this de-associating concept via Example-1 in the first subsections. Now we are able to construct $Y^{\bot}[X_7]$ and all members of $\{X^{\bot}[X_7]_k| k\neq 7\}$ individually and simultaneously lay out their maintained associative relations by conditioning on $X_7$. To do so, we only need to divide the entire data set into $X_7$-specific subsets, denoted ${\cal M}[X_7=j]$ with $J=1, .., 12$. Very importantly, the ${\cal M}[X_7=j]$ retains the associative relations between $Y^{\bot}[X_7]$ and all members of $\{X^{\bot}[X_7]_k| k\neq 7\}$. It is essential to note also that these variables become much less associated with each other. Thus, we can evaluate the effects of all members of $\{X^{\bot}[X_7]_k| k\neq 7\}$ on reducing uncertainty of $Y^{\bot}[X_7]$ beyond the effect of $X_7$.  Since such effects are relatively similar in pattern with respect to different values of $X_7$, we just report such effects in one table with respect to one categorical values of $X_7$.

\begin{table}[h!]
\centering
\begin{tabular}{llllll}\hline
1Feature & CE[$X_7=8$]	&CE[$X_7$] &2Feature   &CE[$X_7=8$]&CE[$X_7$]  \\ \hline
X4&0.7972&0.7648&X3\_X4&0.7350&0.7048\\
X3&1.0827&1.0152&X1\_X4&0.7807&0.7487\\
X5&1.1107&1.0448&X2\_X4&0.7834&0.7486\\
X2&1.1117&1.0419&X4\_X6&0.7850&0.7521\\
X6&1.1125&1.0445&X4\_X8&0.7856&0.7527\\
X1&1.1136&1.0459&X3\_X6&1.0596&0.9924\\
X8&1.1146&1.0473&X3\_X8&1.0631&0.9951\\
X10&1.1147&1.0470&X2\_X3&1.0653&0.9948\\
X9&1.1151&1.0472&X1\_X8&1.0938&1.0253\\
X7&1.1170&1.0498&X7\_X9&1.1151&1.0472\\
\hline
\end{tabular}%
\caption{Example-1 with  $N=10^5$ and CEs when $X_7=8$ and weighted CEs when $X_7$ ranging the 12 categories. The response variable is $Y^{\bot}[X_7]$ and covariate features $\{X^{\bot}[X_7]_k| k\neq 7\}$. }
\label{tableYbyX7eq8}
\end{table}

On the Table~\ref{tableYbyX7eq8}, we can evidently see in the 1-feature setting that $X^{\bot}[X_7]_4$ achieves the lowest CE conditioning on $X_7=8$. Since this evident pattern occurs across all different values of $X_7$. We can conclude that $X_4$ indeed brings extra information beyond $X_7$. That is, $\{X_4, X_7\}$ is confirmed as an order-2 major factor of Re-Co dynamics of $Y$.

As for the feature $X_3$, we observed that the CE of $X^{\bot}[X_7]_3$ is also visible comparing with CEs among $\{X^{\bot}[X_7]_5, X^{\bot}[X_7]_6, X^{\bot}[X_7]_8, X^{\bot}[X_7]_9, X^{\bot}[X_7]_{10}\}$, which are basically noise features after conditioning on $X_7=8$. This observed pattern is also persistent across all different values of $X_7$. Further, since the two feature $\{X^{\bot}[X_7]_3, X^{\bot}[X_7]_4\}$ are no longer highly associated, and this feature-pair $\{X^{\bot}[X_7]_3, X^{\bot}[X_7]_4\}$ achieves a CE in 2-feature setting in Table~\ref{tableYbyX7eq8} that satisfies the ecological effect. Thus, $X_3$ is confirmed as an order-1 major factor.

As for features $X_1$ and $X_2$, their CEs in 1-feature setting are rather close to those perceived noise-features. But the CEs of feature-pairs $\{X^{\bot}_1[X_7], X^{\bot}_4[X_7]\}$ and $\{X^{\bot}_2[X_7], X^{\bot}_4[X_7]\}$ seem to indicate slight effects of these two features, but not evident enough to achieve achieve the ecological effect. That is, it is not confident to conclude whether $X_1$ or $X_2$ will bring extra effects on top of that of $\{X_4, X_7\}$ like $X_3$ does.

To further confirm whether $X_1$, $X_2$ or $X_3$ indeed have extra effects beyond $\{X_4, X_7\}$, we perform likewise de-associating operation ( conditioning) based on bivariate values of $\{X_4, X_7\}$. That is, we look into associative relations between $Y^{\bot}[X_4,X_7]$ and all members of $\{X^{\bot}_k[X_4, X_7]| k\neq 4, 7\}$. It is evident from Table~\ref{tableYbyX4X7eq88}, we see $X_3$ indeed providing extra information beyond what $\{X_4, X_7\}$ can offer on $Y$. In contrast, it seems that we can confirm to some degree that $X_1$ seemingly has a slight amount of effect going beyond that of $\{X_4, X_7\}$ on $Y$ as well. In conclusion, we have the authentic collection of major factors: $\{\{X_4, X_7\}, X_3, X_1\}$, that is, one order-2 and two order-1 major factor for the Re-Co dynamics of $Y$.

\begin{table}[h!]
\centering
\begin{tabular}{llllll}\hline
1Feature &  $CE^{\bot}[X_4, X_7]$ & weighted $CE^{\bot}[X_4, X_7]$	 &2Feature   & $CE^{\bot}[X_4, X_7]$  \\ \hline
X3&0.7441&0.7048&X3\_X10&0.7013\\
X2&0.7978&0.7486&X1\_X3&0.7131\\
X1&0.7931&0.7487&X2\_X3&0.7170\\
X5&0.7978&0.7524&X3\_X6&0.7204\\
X8&0.7982&0.7527&X4\_X8&0.7856\\
X6&0.7992&0.7521&X8\_X9&0.7474\\
X9&0.7993&0.7536&X1\_X8&0.7586\\
X10&0.8002&0.7528&X2\_X8&0.7605\\
X4&0.8032&0.7648&X1\_X2&0.7668\\
X7&0.8032&0.7648&X4\_X7&0.8032\\
\hline
\end{tabular}%
\caption{Experiment-20220304 with  $N=10^5$ and $(X_4,X_7)=(8,8)$ with response $Y^{\bot}[X_4, X_7]$ and covariate features $\{X^{\bot}[X_4, X_7]_k| k\neq 4, 7\}$ at 2nd and 5th column. At 3rd column, weighted CEs of features when conditioning on $\{X_4, X_7\}$ across $(12)^2$ 2D categories.}
\label{tableYbyX4X7eq88}
\end{table}

\subsection{Major factor selection on the Example-2}
First, we recall the structural linearity in the example-2 as follows:
\begin{eqnarray*}
Y&=&0.8 X_1 + X_2 + 1.2 X_3+\epsilon,\\
X_7&=& (X_1+X_2+X_3+X_4+X_{11})/3.66,\\
X_i &\sim& N(0,1), \; \rho(X_i, X_j)=0.7, \; i,j=1,..,6;\\
X_k &\sim& N(0,1), i.i.d,\; k=8,...,11,\\
X_{11}&:& \textrm{is an unobserved hidden variable}.
\end{eqnarray*}
Its 100K simulated data set is created by retaining all data of covariate features in Example-1, but only correspondingly changes the $Y$ values according to its formula. Thus, the the associative heatmap, graph or network among these 12 variables is expected to be very much alike Figure~\ref{fig:20220304MCE}. We compute and report CEs across three feature settings in Table~\ref{table20220408} and summarize our first phase of major factor selection by pointing out major factor candidates across the three feature settings, respectively.
\begin{description}
\item[1-feature setting:] The feature $X_7$ achieves the lowest CE followed by CEs of $X_3$, $X_2$ and $X_1$, in the order of increasing values. They are evident candidates of order-1 major factors. Since their CEs are significantly smaller than CEs of individual features of pure noise $\{X_8, X_9, X_{10}\}$. However, the nearly constant CEs of $\{X_4, X_5, X_{6}\}$ are smaller than CEs of $\{X_8, X_9, X_{10}\}$, but significantly larger than the 4th ranked CE of $X_1$. This observed pattern could be entirely due to their high degrees of dependence with $\{X_1, X_2, X_{3}\}$. Thus, their candidacy for order-1 major factor is not clear and need de-associating operation to confirm or dismiss.
\item[2-feature setting:] The feature-pair $\{X_2, X_3\}$ achieves the lowest CE, not the feature-pair $\{X_3, X_7\}$, which has a CE even larger than CE of feature-pair $\{X_1, X_3\}$. The feature-pair $\{X_2, X_3\}$ seems to be at the position of the feature-pair $\{X_4, X_7\}$ in the previously discussed Example-1. No pairs achieve ecological effect. These results seem to indicate that the role $X_7$ in reducing uncertainty of $Y$ is out-performed by $X_3$. For this reason, we need to perform de-associating with respect to $X_3$ and $X_7$, respectively, to further confirm candidacy of order-1 and order-2 major factors.

    On the other hand, all three feature-pairs of members of $\{X_4, X_5, X_{6}\}$ do not show any significant CE-drops from their individual members. Thus, they are not likely order-2 major factors.

\item[3-feature setting:] The feature-triplet $\{X_1, X_2, X_3\}$ achieves the lowest CE with a significant CE-drop from CE of the feature-pair $\{X_2, X_3\}$. This CE of feature-triplet  $\{X_1, X_2, X_3\}$ is significantly smaller than CE of $\{X_2, X_3, X_7\}$. This fact clearly indicates that the role of $X_1$ with feature-pair $\{X_2, X_3\}$ is much more critical than that of $X_7$ with feature-pair $\{X_2, X_3\}$. This less important role of $X_7$ can be further reflected by the observation of the two feature-triplets, $\{X_2, X_3, X_6\}$ and $\{X_2, X_3, X_7\}$, are relatively close. We need further evidences from de-associating with respect to $\{X_2, X_3\}$ and $\{X_1, X_2, X_3\}$.
\end{description}

\begin{table}[h!]
\centering
\begin{tabular}{llllll}\hline
1Feature & CE	 &2Feature   &CE	  &3Feature	   & CE	  \\ \hline
X7&1.4205&X2\_X3&1.0300&X1\_X2\_X3&0.5034\\
X3&1.5841&X1\_X3&1.1789&X2\_X3\_X7&0.8849\\
X2&1.6807&X3\_X7&1.1842&X2\_X3\_X6&0.9757\\
X1&1.7576&X2\_X7&1.2725&X2\_X3\_X4&0.9797\\
X4&1.9700&X1\_X2&1.3083&X2\_X3\_X9&1.0048\\
X6&1.9705&X1\_X7&1.3320&X1\_X3\_X7&1.0125\\
X5&1.9729&X6\_X7&1.3773&X1\_X3\_X4&1.1177\\
X10&2.4103&X4\_X7&1.4154&X1\_X3\_X6&1.1188\\
X8&2.4104&X4\_X5&1.8132&X1\_X2\_X7&1.1283\\
X9&2.4104&X8\_X9&2.4030&X8\_X9\_X10&2.3080\\
\hline
\end{tabular}%
\caption{Experiment-20220408 with  $N=10^5$. Each categorized 1-features has 12 bins, so a $k$-feature setting has $(12)^k$ $k$D hypercubes.}
\label{table20220408}
\end{table}

Next we implement de-associating computations with respect to $X_7$ first and $X_3$ secondly, and report CEs of all $X^{\bot}[X_7]_i$ and $X^{\bot}[X_3]_i$ correspondingly in Table ~\ref{20220408botbyX7} and Table~\ref{20220408botbyX3}. It is worth noting that members of either $\{X^{\bot}[X_7]_i\}$ or $\{X^{\bot}[X_3]_i\}$ are much less associated with each other. Results pertaining to de-associating with respect to $\{X_2, X_3\}$ and $\{X_1, X_2, X_3\}$ are reported in Table~\ref{20220408botbyX3X2X1}. We summarize results from these three tables to advance our major factor selection protocol for Example-2.
\begin{description}
\item[[Table ~\ref{20220408botbyX7}]:] From its 3rd and 6th columns, we see the three pairs: $\{X^{\bot}[X_7]_2, X^{\bot}[X_7]_3\}$, $\{X^{\bot}[X_7]_1, X^{\bot}[X_7]_3\}$ and $\{X^{\bot}[X_7]_1, X^{\bot}[X_7]_2\}$, show their ecological effects, simultaneously. So they can be concurrently order-1 major factors, but they are not potential order-2 major factors candidates. The pair $\{X^{\bot}[X_7]_3, X^{\bot}[X_7]_6\}$ does not have ecological effect.

\item[[Table~\ref{20220408botbyX3}]:] From its 3rd and 6th columns, we see pair $\{X^{\bot}[X_3]_1, X^{\bot}[X_3]_2\}$ satisfies the ecological effect. Neither $\{X^{\bot}[X_3]_2, X^{\bot}[X_3]_7\}$,  nor $\{X^{\bot}[X_3]_2, X^{\bot}[X_3]_6\}$ show ecological effects.

\item[[Table~\ref{20220408botbyX3X2X1}]:] When performing de-associating with respect to $\{X_2, X_3\}$, we see that $X_1$ achieves the lowest weighted CE across all $(12)^2$ categories, which is significantly lower than that of $X_7$. That is, $X_1$ still provides an extra amount of uncertainty reduction on $Y$ beyond $\{X_2, X_3\}$. Though, $X_7$ also provides a small amount of uncertainty reduction on $Y$ beyond $\{X_2, X_3\}$ as expected. Further, when performing de-associating with respect to $\{X_1, X_2, X_3\}$, we see that $X_7$ achieves a CE even larger than the pure noise features $X_8$, $X_9$ or $X_{10}$. That is, $X_7$ indeed does not provide any extra information beyond what $\{X_1, X_2, X_3\}$ can provide. Likewise, individual feature like $X_4$, $X_5$ or $X_6$ do not offer any extra information beyond what $\{X_1, X_2, X_3\}$ can offer.

\item[[Conclusion:]] It is evident that $\{X_1, X_2, X_3\}$ is a collection of three order-1 major factors. In contrast, feature-pair $\{X_4, X_7\}$ is not alternative collection of order-2 major factors. And $X_5$ and $X_6$ do not have any roles as order-1 major factor.

\end{description}

\begin{table}[h!]
\centering
\begin{tabular}{llllll}\hline
1Feature & CE[$X_7=8$]	&CE[$X_7$] &2Feature   &CE[$X_7=8$]&CE[$X_7$]  \\ \hline
X3&1.2598&1.1842&X2\_X3&1.0787&0.8849\\
X2&1.3611&1.2725&X1\_X3&1.2111&1.0125\\
X1&1.4255&1.3320&X1\_X2&1.2162&1.1283\\
X6&1.4726&1.3773&X3\_X6&1.2170&1.1390\\
X5&1.4734&1.3782&X3\_X8&1.2404&1.1596\\
X4&1.5085&1.4154&X3\_X7&1.3137&1.1842\\
X10&1.5090&1.4161&X2\_X8&1.3350&1.2457\\
X9&1.5099&1.4165&X3\_X4&1.2598&1.3481\\
X8&1.5090&1.4161&X4\_X7&1.5090&1.4154\\
X7&1.5130&1.4205&X7\_X9&1.5099&1.4165\\
\hline
\end{tabular}%
\caption{Experiment-20220408 with  $N=10^5$ and CEs when $X_7=8$ and weighted CEs when $X_7$ ranging the 12 categories. The response variable is $Y^{\bot}[X_7]$ and covariate features $\{X^{\bot}[X_7]_k| k\neq 7\}$. }
\label{20220408botbyX7}
\end{table}

\begin{table}[h!]
\centering
\begin{tabular}{llllll}\hline
1Feature & CE[$X_3=8$]	&CE[$X_3$] &2Feature   &CE[$X_3=8$]&CE[$X_3$]  \\ \hline
X2&1.0970&1.0300&X1\_X2&0.5280&0.5034\\
X1&1.2505&1.1789&X2\_X7&0.9460&0.8849\\
X7&1.2587&1.1842&X2\_X6&1.0436&0.9757\\
X4&1.5468&1.4617&X2\_X4&1.0469&0.9797\\
X6&1.5500&1.4600&X2\_X8&1.0724&1.0046\\
X5&1.5532&1.4616&X1\_X7&1.0735&1.0125\\
X10&1.6717&1.5793&X1\_X6&1.1876&1.1188\\
X8&1.6721&1.5791&X1\_X8&1.2243&1.1488\\
X9&1.6724&1.5793&X4\_X7&1.2403&1.1651\\
X3&1.6766&1.5841&X3\_X9&1.6724&1.5791\\
\hline
\end{tabular}%
\caption{Experiment-20220408 with  $N=10^5$ and CEs when $X_3=8$ and weighted CEs when $X_3$ ranging the 12 categories. The response variable is $Y^{\bot}[X_3]$ and covariate features $\{X^{\bot}[X_3]_k| k\neq 3\}$. }
\label{20220408botbyX3}
\end{table}

\begin{table}[h!]
\centering
\begin{tabular}{lll}\hline
1Feature & $CE^{\bot}[X_2, X_3]$	&$CE^{\bot}[X_1, X_2, X_3]$ \\ \hline
X1&0.5034&0.5034\\
X2&1.0300&0.5034\\
X3&1.0300&0.5034\\
X4&0.9797&0.4518\\
X5&0.9772&0.4500\\
X6&0.9758&0.4508\\
X7&0.8849&0.4686\\
X8&1.0046&0.4359\\
X9&1.0048&0.4356\\
X10&1.0043&0.4344\\
\hline
\end{tabular}%
\caption{Experiment-20220408 with  $N=10^5$ and weighted CEs conditioning on (boted by)$\{X_2, X_3\}$ and weighted CEs when conditioning on (boted by) $\{X_1, X_2, X_3\}$ across all 2D and 3D categories. The response variable are $Y^{\bot}[X_2, X_3]$ and $Y^{\bot}[X1, X_2, X_3]$, and covariate features $\{X^{\bot}[X_2,X_3]_k| k\neq 3\}$ and $\{X^{\bot}[X_1, X_2,X_3]_k| k\neq 3\}$, respectively. }
\label{20220408botbyX3X2X1}
\end{table}

\subsection{Major factor selection (MFS) protocol}
Through the Example-2, we fully illustrate our major factor selection protocol as an operational process. In this subsection, we lay out a step-by-step operational process for this protocol. This description is intended to make our major factor selection protocol an highly adaptable way of studying Re-Co dynamics embedded within any structured data sets derived from complex systems of interest.
\begin{description}
\item[MFS-1:] Based on a matrix representation of an observed structured data set, we first explore potentially structural associations among all features by using a mutual conditional entropy (MCE) heatmap superimposed with a hierarchical clustering tree. Various compositions of block patterns in MCE heatmap reveal various maps of community-based structural dependency across all involving features on both response and covariate sides. The response feature-set is denoted as ${\cal Y}$.

\item[MFS-2:] To a great extent, such structural  dependency, which maps out which features being highly associated with which features, but not so much with other features, will help explain and make senses of all calculated CEs of all feature-sets across multiple feature-settings. Based on explained CEs and corresponding CE-drops (conditional mutual information), we identify highly potential candidates of major factors of low orders: either order-1 or order-2, that are able to achieve significant CE-drops. Denote these low-order major factor candidates as $\{A_1, A_2, .., A_K\}$ with $K$ being a suitable number, which is not necessarily equal to the notation total number of covariate features with the same notation.

\item[MFS-3:] We them perform the de-associating operations with respect to individual $A_k$ with $k=1,.., K$ to confirm its candidacy and simultaneously further discover which members of $\{A_1, A_2, .., A_K\}$ or any feature-sets outside of this collection that indeed offer significant CE-drops with respect to dynamics of ${\cal Y}^{\bot}[A_k]$ across all categorical values of $A_k$. This discovery is carried out with helps based on the two criterions proposed in \cite{CCF22a,CCF22b}. These two independence-based criterions become much more relevant and suitable because covariate features are significantly less associative after the de-associating operation.
 \begin{description}
\item[[C1: Confirmable]:] A feature-set $B$ is confirmable if a feature-set $\tilde{B}$ is obtained by substituting anyone of feature members of $B$ with a feature that is completely independent of ${\cal Y}^{\bot}[A_k]$ and $B$, we have $I[{\cal Y}^{\bot}[A_k];]$  is significantly larger than $I[{\cal Y}^{\bot}[A_k];\tilde{B}]$.
\item[[C2: unreplaceable]:] A feature-subset $B$ is replaceable if $I[{\cal Y}^{\bot}[A_k];B] \leq  I[{\cal Y}^{\bot}[A_k];B_1]+I[{\cal Y}^{\bot}[A_k];B_2]$ for any compositions of $B$, i.e. $B= B_1\bigcup B_2$ and $B_1\bigcap B_2=\emptyset$. For $B$ to be declared as unreplaceable, we require that $B$ is not replaceable and simultaneously satisfies the following two extra conditions: (a) its CE-drop is larger the sum of the top ranked CE-drop and at least $|B|$-times of its complementary feature-subset' CE-drop; (b) the candidate $B$ joins with any already identified major factor $B^*_m$ must achieve $I[{\cal Y}^{\bot}[A_k];B\bigcup B^*_m] \geq  I[{\cal Y}^{\bot}[A_k];B]+I[{\cal Y};B^*_m]$.
\end{description}

\item[MFS-4:] The major factor selection results in MFS-3 could vary among all categorical values of $A_k$. This phenomenon of having varying multiscale major factor reveals the structural heterogeneity in the dynamics of ${\cal Y}$ as well as in data's information content. If some $k'\neq k$ and $A_{k'}\in \{A_1, A_2, .., A_K\}$ never get selected, then apparently $A_{k'}$ doesn't offer information of ${\cal Y}$ beyond $A_k$. As such the major factor candidacy of $A_{k'}$ is revoked. On the other hand, if $A_{k'}$ uniformly induced significant CE-drops across all categories of $A_{k}$, then not only major factor candidacy of $A_{k'}$ is reconfirmed, but also bring out the necessity of performing de-associating with respect to $(A_k, A_{k'})$, that is, we need to look into dynamics of ${\cal Y}^{\bot}[A_k, A_{k'}]$ and repeat the MFS-3 step.
\end{description}
It is noted that, the criterion [C1: Confirmable]  in MFS-3 is mainly used as a reliability check, while the criterion [C2:irreplacable] focuses on checking the fact about whether feature-sets can offers information beyond a designated major factor candidate, which is termed ecological effect of these two feature-sets. When the ecological effect is found significantly high between two feature-sets, this criterion provides an effective tool to confirm the discovery of their interacting effect. This is the most essential way of discovering a high order major factor as a composition of marginally very less associated covariate features that become highly dependent given ${\cal Y}$ . The various potential scenarios in MFS-4 are already seen in Example-1 and Example-2.
 
After our computational developments and illustrations through Example-1 and Example-2, the above major factor selection protocol is proposed with helps from structural dependency among all features to bring out structural heterogeneity based multiscale information content. Such a protocol is particularly needed for analyzing structured data sets derived from large complex systems. We would witness clearly how this protocol works out in two real-world applications reported in the next two sections.

\section{Heart disease's complex dynamics.}
The Behavioral Risk Factor Surveillance System (BRFSS), established in 1981 \cite{remington}, is a collaborative project between all of the states in the United States and participating US territories and the Centers for Disease Control and Prevention (CDC). BRFSS is a health-related telephone survey collected annually by the CDC. Each year, the survey collects responses from over 400,000 Americans on health-related risk behaviors, chronic health conditions, and the use of preventative services. BRFSS's objective as stated in its website is `` to collect uniform state-specific data on health risk behaviors, chronic diseases and conditions, access to health care, and use of preventive health services related to the leading causes of death and disability in the United States.''

We analyze a cleaned and consolidated data set created from the Behavioral Risk Factor Surveillance System (BRFSS) 2015 dataset on Kaggle (
\textrm{https://www.kaggle.com/alexteboul/heart-disease-health-indicators-dataset}). This data set contains 253,680 survey responses, in which 229,787 respondents do not have/have not had heart disease while 23,893 have had heart disease. With an almost 10-to-1 ratio of class imbalance, this data set is used primarily for the binary classification of heart disease aiming at exploring the following two questions: Q1:)To what extend can survey responses from the BRFSS be used for predicting heart disease risk? Q2:) Can a subset of questions from the BRFSS be used for preventative health screening for diseases like heart disease?

When applying many off-shelf machine learning approaches, such as logistic regression, Random Forest and various Boosting, all the accuracies are not better than $90\%$. The easily identified primary cause of such a phenomenon is the non-diseased-vs-diseased imbalance: $(0.904, 0.096)$. That is, we can blindly adopt the constant predictive rule: ``all subjects are non-diseased'', to have an accuracy slightly better than $90\%$. Nonetheless, such an extreme imbalance is natural and ubiquitous across all chronic diseases. 

So the data analysis challenge is immanent. The BRFSS survey population (18 year old and beyond) is large and diverse. In the US, the heart disease (HD) has been one of the leading cause of death for several decades. It annually claimed more than 600 thousands life. Unfortunately,  even up to now, the answers to the above two questions are not well established yet. Here we hope our CEDA-based resolutions, as would be given below, can carry some great impacts to human societies on the regard of heart disease and beyond.

According to BRFSS Data Codebook, MICHD is a calculated variable for respondents that have ever reported having coronary heart disease (CHD) or myocardial infarction (MI). Myocardial infarction is also commonly termed as Heart Attack.  In this paper, we adopt the variable names used in the Kaggle version of data set. So MICHD is replaced by HDAtt and is taken as the binary response-feature of the targeted Re-Co dynamics. All other variables are taken as covariate features in this Re-Co dynamics. Each covariate feature's abbreviated names is defined at its first appearance in the text. Further, since HDAtt is binary, we present and display results of our selected major factors in terms of odds, which is defined as the proportion of diseased against the proportion of non-diseased within each category of a covariate feature-set. One odds corresponds to one CE.

\begin{figure}
 \centering
  \includegraphics[width=6in]{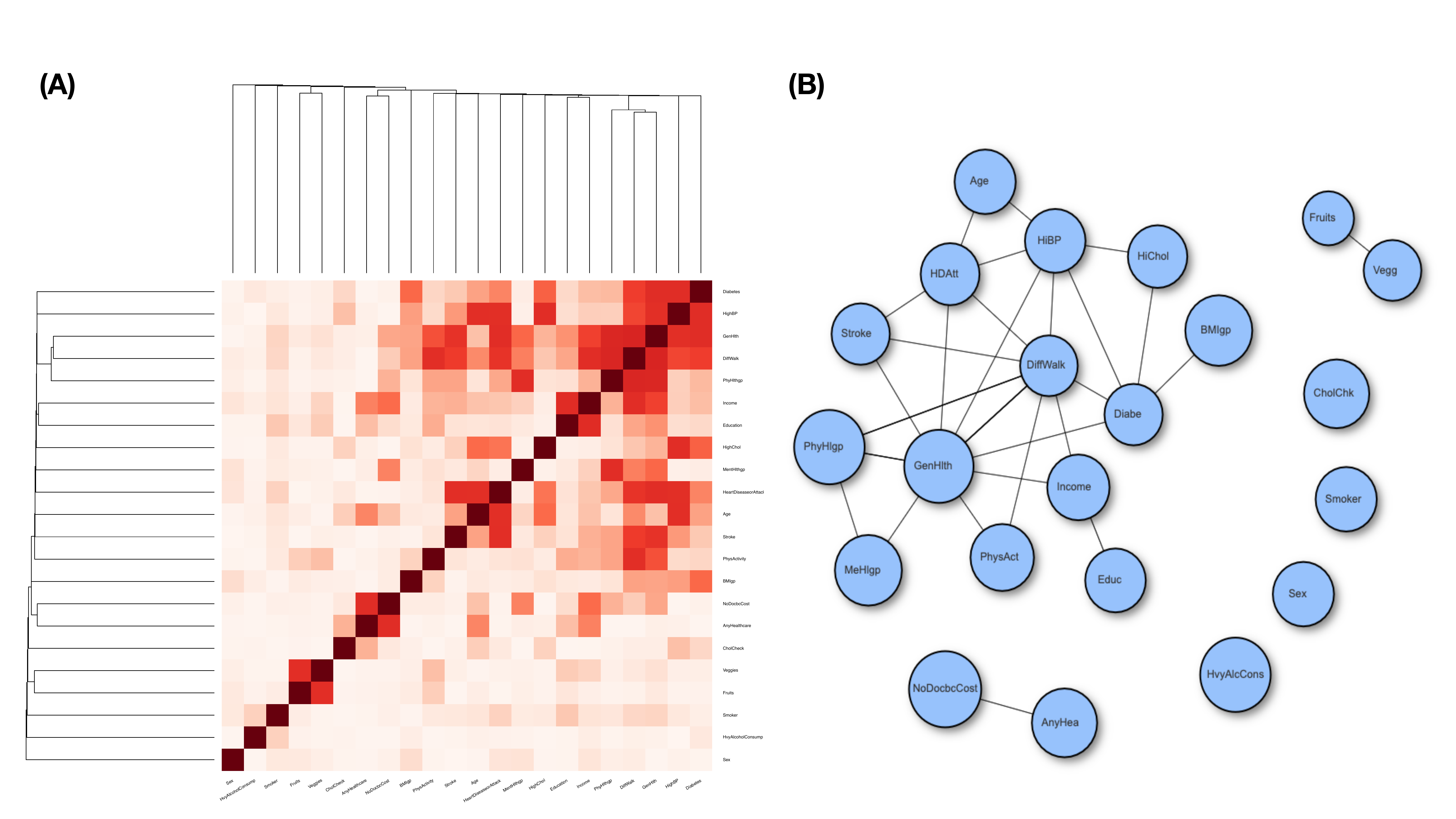}
 \caption{MCE Heatmap of all features in Kaggle's HD data.}
 \label{MCEKaggle}
 \end{figure}

Upon studying this targeted Re-Co dynamics of HDAtt, a glimpse of challenging aspect is immediately seen from the MCE-based heatmap of all involving features in Figure~\ref{MCEKaggle}. By carrying out the MFS-1 step, we see complicate and structural associative patterns among all involving features. In particular, the response feature HDAtt is highly associated with features: Age and GenHl (general Health), and mildly associated with many others.

Another challenging aspect of this data can be seen through a contingency table in Table~\ref{CholHD} coupled with low-wise diseased-vs-non-diseased proportion. As a single binary feature, high Cholesterol (HiChol) achieves an odds-ratio $3.6= \frac{16753}{90838}/\frac{7140}{138949}$, while another binary feature high Blood Pressue (HiBP) achieves an even higher odds-ratio $\frac{17928}{90901}/\frac{5965}{138886}=\frac{0.197}{0.043}=4.5.$. Though, the two odds-ratios can successfully convey the relative risks of being in the category of $HiChol=1$ or $HiBP=1$, the diseased ones are still minorities in both categories. This phenomenon of being minority is also clearly seen in the four bivariate categories of (HiBP, HiChol) in Table~\ref{CholHD}. Nonetheless, the odds-ratio measure doesn't well suit for the setting having four bivariate categories. However, the phenomenon of the diseased minority hiding behind the non-diseased majority is still persistently clear.

\begin{table}[h!]
\centering
\begin{tabular}{llll}\hline
HiBP-HiChol/HD & non-diseased & diseased & prob-vector\\ \hline
0-0&99044&2876&(0.972, 0.028)\\
0-1&39842&3089&(0.929, 0.071)\\
1-0&39905&4264&(0.904, 0.096)\\
1-1&50996&13664&(0.733, 0.267)\\\hline
C-sums&229787 &23893 &(0.906, 0.094)\\
\hline
\end{tabular}%
\caption{Contingency table of (HiBP, HiChol)-vs-HDAtt.}
\label{CholHD}
\end{table}

This persistent and wide-spreading minority-hiding-behind-majority phenomenon surely sheds lights on the complexity of this disease dynamics. In this section, to a great extent, we break this class-imbalance based phenomenon by applying our major factor selection protocol to manifest the data's informative structural heterogeneity. That is, we would be able to peek into this complex dynamic system of heart disease by mapping out such profound structural heterogeneity embraced in this BRFSS data set by applying our major factor selection protocol.

\begin{figure}
 \centering
  \includegraphics[width=5in]{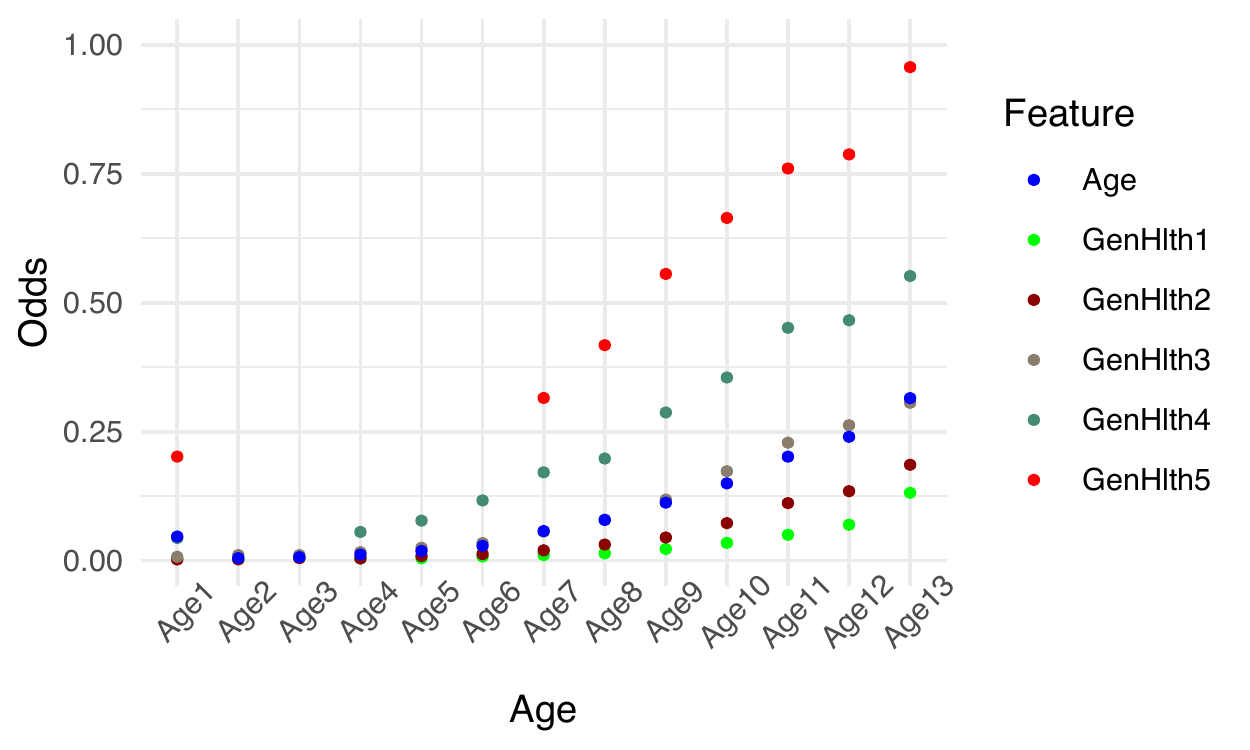}
 \caption{Odds evolutions w.r.t features.}
 \label{odds1}
 \end{figure}

On the MFS-2 step of major factor selection protocol, we discover that Age and General Health (GenHl) are two dominate order-1 major factors. We present such findings through the expansions of odds with respect to the 5 categories of GenHl across 13 age categories in Fig~\ref{odds1}. The odds (blue dots) of 12 age-categories, except the 1st one, are evidently increasing as ages increasing. Upon each age category by centering GenHl=3, particularly from age-7 to age-13, the 2 odds of GenHl expand upward for GenHl=4 and GenHl=5, while the 2 odds for GenHl=1 and GenHl=2 go downward. Such expansions are wider and wider as age increasing. In this fashion, these 65 localities constitute the large scale of structural heterogeneity embedded within the Re-Co dynamics of HDAtt.

We then proceed to the MFS-3 and MFS-4 steps of our major factor selection protocol to work on each of the 65 localities resulted from the MFS-2. In this study, the feature Diabetes (Diabe) is made into a binary feature by merging the relatively very small category 1 with category 2. Within each locality with sample sizes over 500, we select a triplet of binary features that achieve the largest CE-drop and report our selected triplet of order-1 major factors in Table~\ref{triplets}. These triplets provide the largest amount of information on HDAtt status beyond the Age and GenHl. For those localities with sample sizes less than 500, a NA is reported.

\begin{figure}
 \centering
   \includegraphics[width=6in]{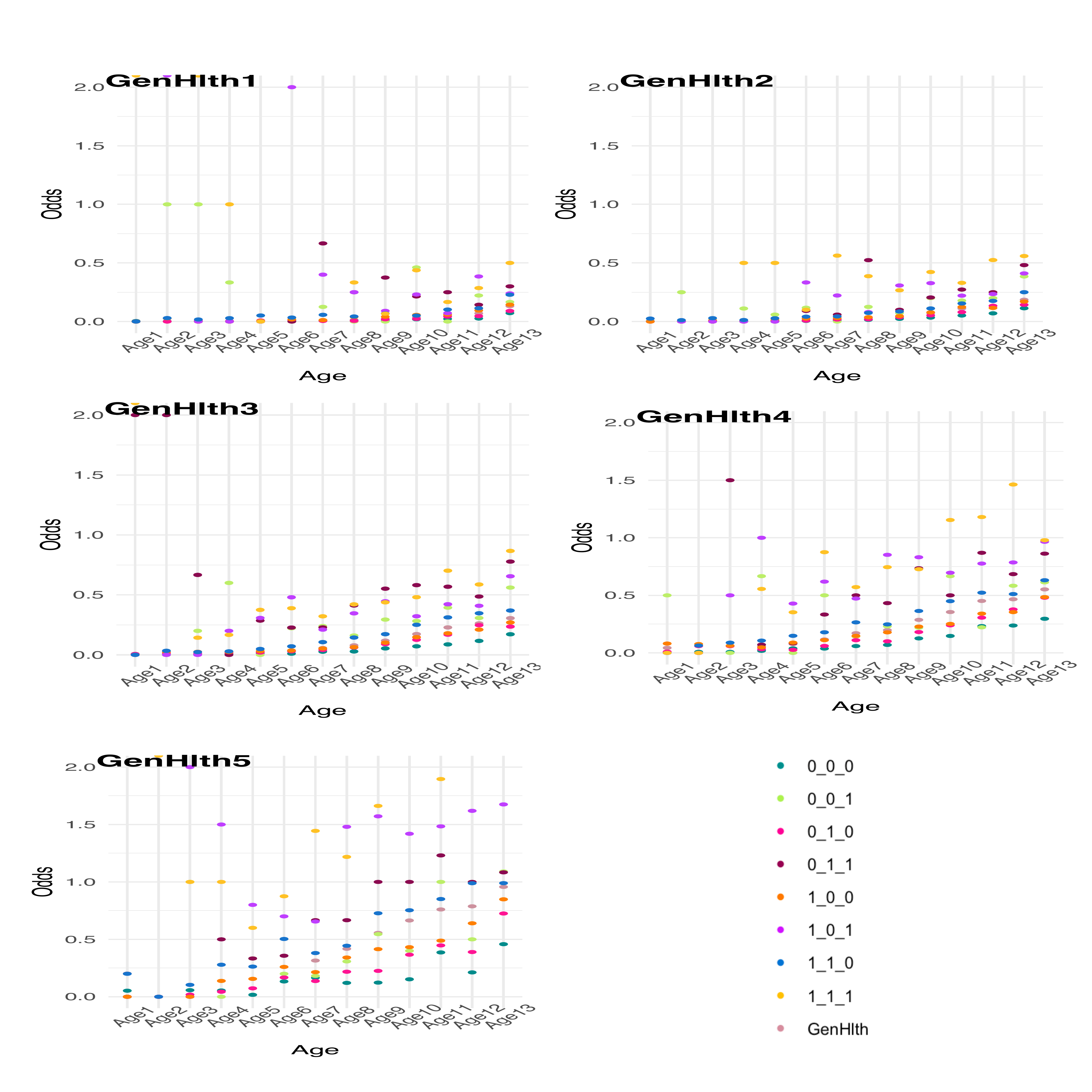}
 \caption{Odds-expansions w.r.t selected triplet features upon 48 localities defined based on (Age, GenHl).}
 \label{odds2}
 \end{figure}

Upon each selected triplet at a locality, we report the 8 odds and show their clear odds-expansions. There are 48 localities are reported in Fig~\ref{odds2} with respect to GenHl's 5 categories. Through the 5 GenHl-specific panels, the global pattern of growing odds-expansions is evidently seen. Within each panel, we visualize another growing pattern of odds-expansions across the across age axis. In particular, many odds in the 5th panel (GenHl=5) are greater than 1. That is,within each of such localities, the diseased has become the majority. In fact, this figure fully displays all localities with elevated odds implying that probabilities of having HD becomes dangerously real. 

From this perspective, Fig~\ref{odds2} indeed provides exquisite information content of this data set. Its merits and implications are significant. Since a medical doctor not only is able to precisely assess his/her patients' chances of having heart disease with respect to patient's locality, but also is able to persuade patients to stay away from the direction of the expanded odds.This kind of assessment realistically amounts to be much more powerful than those based on odds-ratios. Finally, as a byproduct of predictive inferences, the simply majority-rule would do much better than the blind-rule and all aforementioned machine learning approaches. In this fashion, we achieve the positive resolutions for the aforementioned two questions at the beginning of this section. We hope that these resolutions through data's information content represented in Fig~\ref{odds2}bring insightful understanding on behavioral risks to this heart disease, and likewise for other chronic diseases.

\begin{table}[]
\resizebox{\textwidth}{!}{%
\begin{tabular}{llllllllllllll}\hline
GenHL/Age & 1 & 2 & 3 & 4 & 5 & 6& 7& 8& 9& 10& 11& 12& 13 \\ \hline
1 & FSV & FSV& DSV & BSV& FOW& CDF& CDS& BDS& BWX& CDS&  BDS&  BCX& BFX\\
2 & BFO & BFS& BCF & CSX & BDO& BSX& BCD& BCD& BCX&  CSX&  BCX&  CSX& CSX\\
3&OSX & BOS& BCS& BOS& CSW&  BCS&  BCS&  BCS&  BCS& CSX&  CSX&   CSX&  CSX\\
4 &NA&   NA&   BOS& BOS& BDW& BCS&  BCS&  BCS&  CSX&  CSX& CSX&  CSX&  CSX\\
5 &NA&   NA&    NA&  NA&    NA&    CDO& BOS& DSX&  BDS&  DOS& DSX& BDS& BDS\\
\hline
\end{tabular}%
}
\caption{The selected triplets with the response variable is $Y=HD^{\bot}[Age, GenHl]$ and covariate features $\{X^{\bot}[Age, GenHl]_k\}$. B=HBP; C=highChol; S=Stroke; D=Diabetes; X=Sex; O=Smoker; W=DiffWalk; F=Fruits; V=Veggies.}
\label{triplets}
\end{table}

\section{MLB Fastball pitching dynamics}
Beyond being a sport, the baseball pitching is a well-known example of Magnus effect from physics perspective \cite{briggs}. This effect describes how one spinning object would be subject to a spin-generated force. This force is perpendicular to the spinning object's traveling direction and in principle will make it curves according to spin direction and rate. For example, the spin direction of fastball pitch is primary back-spinning. Its Magnus effect will cause the fastball to curve upward, looking like being pushed up against the gravity. Fastball is the most dominant pitch type in Major League Baseball (MLB) in the US. See more detail descriptions and discussions regrading Magnus effects on other pitch-types in  \cite{FC21,FCC21,CCF22a,CCF22b}.

Each professional baseball pitcher in MLB has his own unique way of creating fastball-oriented biomechanical forces by coupling with varying versions of back-spins. His pitching gesture uniquely characterizes his way of applying his musculoskeletal system to create his biomechanics and specifying the coordinates of baseball's releasing point. And his way of holding his baseball determines a pitch's spin rate and direction. This action is usually hidden within his baseball glove to prevent the batter knowing how his pitch would curve. Further, weather conditions would also complicate a baseball trajectory in a unpredictable fashion, for instance, the precipitation could affect a pitcher's grasp on baseball's leather surface. As such each MLB pitcher has an idiosyncratic pitching dynamics underlying his fastball complex system.

Furthermore, MLB's regular season starts from beginning of April and ends at October. The end of regular season is followed by play-off games that could go into November. Even for a healthy pitcher, his pitching dynamics realistically would more or less evolve in some unknown complicated ways along the course of more than 6 months. Finally, from the perspective of multiple seasons, any professional MLB pitcher's pitching dynamics would go through a certain difficult-to-describe evolution. Therefore, a MLB pitcher's pitching dynamics going over multiple seasons should be taken as a neither well-known, nor well-controlled complex system. Further, a collective system consisting many different pitchers of same pitch-types would be even more complicated because of structural heterogeneity.

\begin{table}[h!]
\centering
\begin{tabular}{llll}\hline
pitcher-name (handed)/season-pitches & 2017 & 2018 & 2019 \\ \hline
Justin Verlander(R)& 2432 & 2276 &1876 \\
Max Scherzer (R) &1589 &1813 & 1596 \\
Chris Archer(R)& 1607 & 951 & 833 \\
Gerrit Cole(R)& 1543 & 1742 & 2041\\
Charlie Morton(R)& 367 & 849 & 1017\\
Jacob deGrom(R) & 1251 & 1469 &1592 \\
Clayton Kershaw(L)& 1385 & 1194 & 1170\\
Chris Sale(L)& 1350 & 1102 & 908\\
Jon Lester(L)& 1336 & 1484 & 910 \\
Mattew Boyd(L)& 619 & 1008 & 1586 \\
Patric Corbin(L)& 840 & 629 & 766\\
Jose Quintana(L)& 1293 & 1480 & 993\\
\hline
\end{tabular}%
\caption{12 fastball pitchers and numbers of pitches across 2017-2019 seasons. (R) for right-handed and (L) for left-handed. }
\label{names}
\end{table}

\begin{figure}
 \centering
  \includegraphics[width=6in]{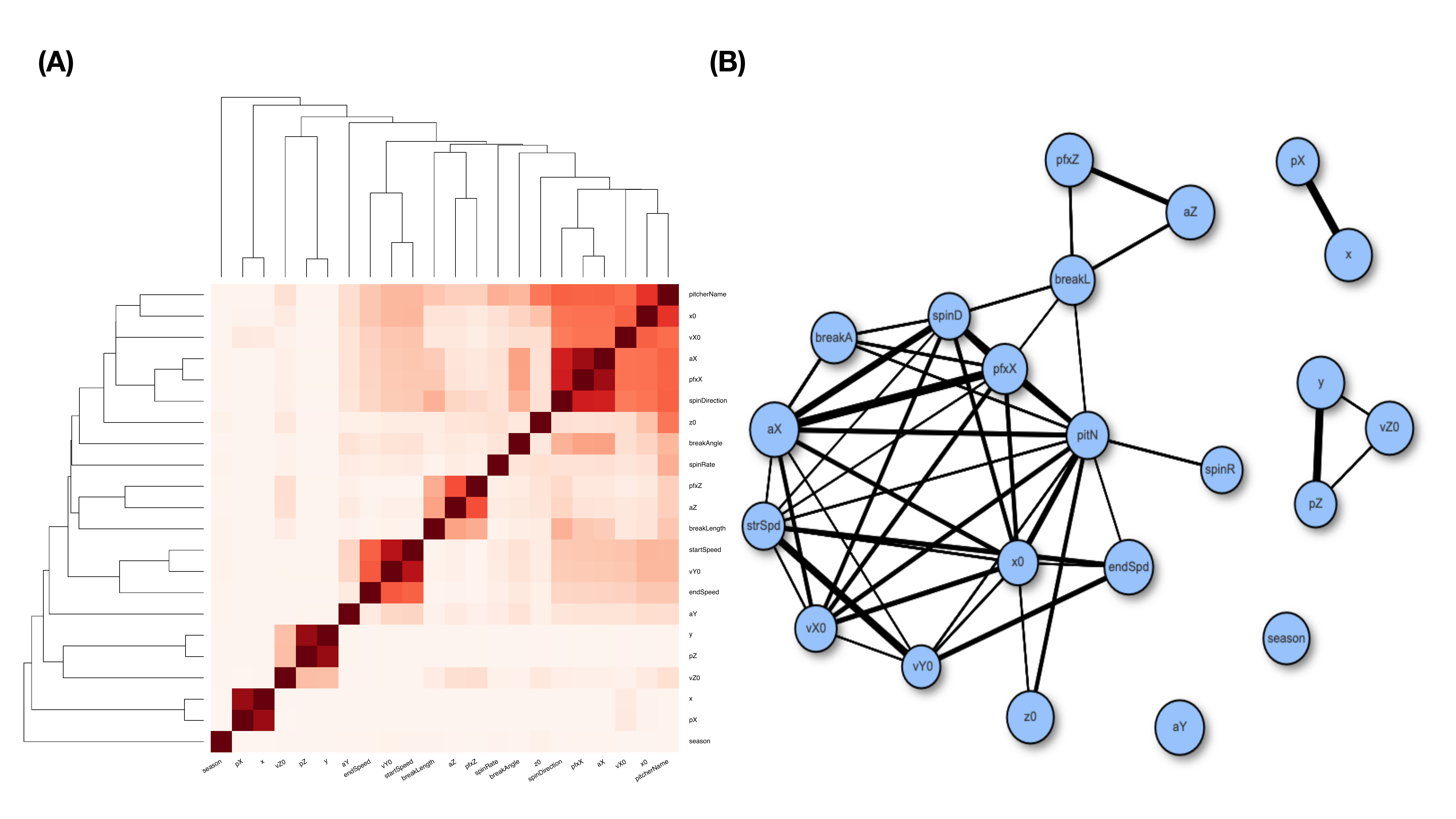}
 \caption{MCE Heatmap and network of all features in MLB's fastball data.}
 \label{MCEMLB}
 \end{figure}

We collected data of 6 right-handed and 6 left-handed pitchers, who pitched the top-6 largest numbers of fastball across 2017 to 2019 seasons from MLB's Statcast. The names of these 12 pitchers are listed in Table ~\ref{names} together with their numbers of fastball pitches across the three seasons. As aforementioned in Introduction section, there are 21 features directly linked to pitchers' pitching dynamics. The names of these features can be seen in Figure~\ref{MCEMLB}, while their detailed descriptions can be found in \cite{FC21,FCC21}. Together with pitcher-name (pitN) and Season as two extra features, we carry out the MFS-1 step of our major factor selection protocol by reporting the MCE-based heatmap and network in two panels of Figure~\ref{MCEMLB}, respectively. Upon the heatmap and network, we see block-patterns and community-revealing linkages collectively indicating highly structural dependency among all features. In particular, we see the two added features: pitN and season, are associative with many features located in different blocks and communities. This empirical fact brings out multiple potential challenges when we further implement our major factor selection protocol.

With each fastball pitch being recorded and described by a 21 dimensional vector labeled by pitcher-name and season, our scientific goal is to explicitly manifest data's information content regarding this complex system of 12 MLB pitchers' collective pitching dynamics. This goal is oriented at least to embrace two chief questions : Q3:) Which part of data's information content will allow us to visualize and discover global and fine scales similarity and difference among these 12 individual complex systems? Q4:) Which part of data's information content will shed lights on fine scale potential changes within a single pitcher's pitching dynamics across the three seasons? Specifically speaking, the Q3 obviously contains the major machine learning topic called MultiClass Classification (MCC), while Q4 clearly attempts to discover unspecified fine scale dynamic changes along a targeted complex systems' evolutions. By addressing Q3 and Q4 in an cohesive fashion here, we want to illustrate an intuitive scientific fact that questions pertaining to a large complex dynamic system are likely connected, so are their resolutions. Here, we stipulate that such connections must chiefly reside in data's authentic information content.

Any real-world large complex system likely contains multi-scale structures characterized by heterogeneous pattern-information, as emphasized in by Nobel physicist P. W. Anderson \cite{anderson}, data derived from such a complex system will likely embraces multiscale information content with heterogeneity. Therefore, data analysts and scientists need to look at questions and to search for resolutions from both multiscale and heterogeneity perspectives. {\bf This idea of ``data's information first and question's resolution second'' might be what the term ``data-driven'' truly means.}

From the multiscale and heterogeneity perspectives, the Q3 indeed embraces that, when each label (pitcher-name) of MCC is commonly defined by a principle physical dynamics and tuned with idiosyncratic characteristics, the MCC resolution can only be efficiently found by firstly pertinently investigating ``the largest common factor'' underlying all labels' dynamics and then ``subtracting this common factor from each individual dynamics'' in order to discover individual-specific characteristics. The piece of information of the largest common factor and all pieces of information of all pitchers' individual dynamics belong to the whole of the data's information content. As for the Q4, it embraces that, even when a question of interest is specifically defined by a fine scale issue within a large complex system, the data's information content involving with many other scales is again needed in order to arrive at a specific fine scale resolution. Indeed the Q1 and Q2 also echo this line of data-driven message.

\subsection{Categorizing Re-Co dynamics and its global information.}
\begin{figure}
 \centering
  \includegraphics[width=6in]{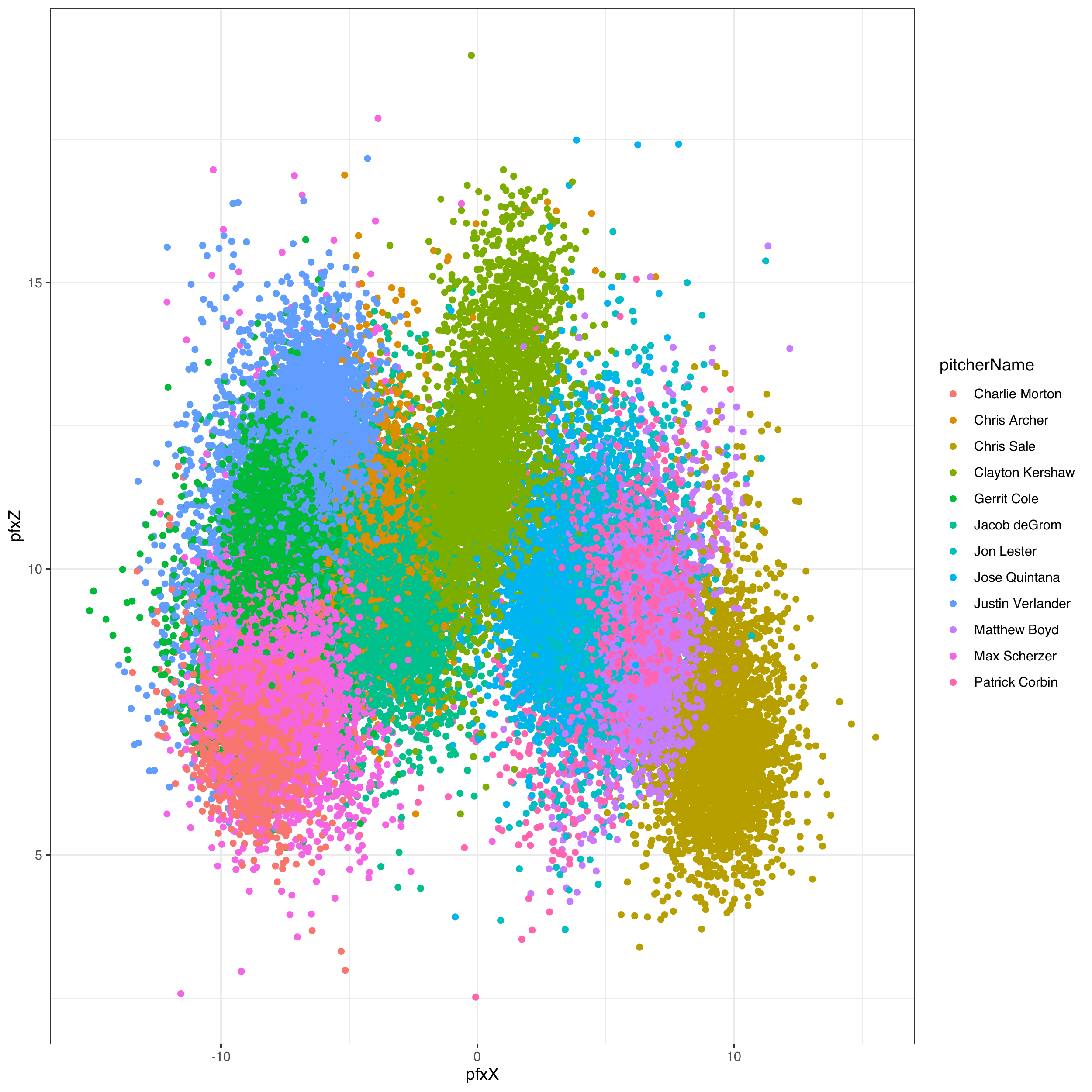}
 \caption{Scatter plot of ${\cal Y}=(pfx_x, pfx_z)$ color-coded with pitcher-name.}
 \label{plot12}
 \end{figure}

We make use of the Re-Co dynamics underlying the 2D response variable ${\cal Y}=(pfx_x, pfx_z)$ to represent the pitching dynamics commonly shared by 12 pitchers. Since ${\cal Y}$, a pitch's horizontal and vertical movements, simultaneously has the two ultimate features that any pitcher wants to deliver when facing a batter at the home plate. This Re-Co dynamics surely provides the global scale information content of these 12 fastball pitchers, see the scatter plot of ${\cal Y}=(pfx_x, pfx_z)$ in Figure~\ref{plot12}.

We then categorize ${\cal Y}$ by taking each occupied cell of $5\times 5$ contingency table $C[pfx_x-vs-pfx_z]$ as one category. That is, $pfx_x$ and $pfx_z$ are categorized via their individual histograms with 5 bins, respectively. So a category of ${\cal Y}$ is a rectangle-cell, which is intuitively more proper than irregular-cell resulted from other categorization schemes, such as using K-mean or hierarchical clustering algorithms. Further, the rectangle-cell would be natural for predictive purpose as well.

With the categorized ${\cal Y}$ and all covariate features, we carry out the MFS-2 by performing the conditional entropy calculations to look into the major factors underlying the Re-Co dynamics. We report the top five feature-sets that achieving the smallest CE across 3 feature settings. On the 1-feature setting, we see five potential candidates of order-1 major factors: 1) $aX$ (horizontal acceleration); 2) $spinD$ (spin direction); 3) $pitN$ ( pitcher IDs); 4) $aZ$ (vertical acceleration); 5) $x0$ (x-coordinate of leasing point). On the 2-feature setting, we see that the feature-pair $(aZ, aX)$ achieves the smallest CE. But due to dependence between $aX$ and $aZ$, this feature-pair's CE-drop is slightly less than the sum of $aX$ and $aZ$'s individual CE-drops, that is, the ecological effect is not observed. However, when we perform the de-associating calculations with respect to $aX$, we see that $aZ$ achieves the smallest CE across all 5 categories of $aX$. That is, $aZ$ certainly contributes extra information on ${\cal Y}$ beyond $aX$. On the 3-feature setting, we do not see significant CE-drops. Therefore, we take the collection $\{aX, aZ\}$ as two separate order-1 major factors. That is, $\{aX, aZ\}$ provides the most significant information content of ${\cal Y}$ on the global scale.

\begin{table}[h!]
\centering
\begin{tabular}{llllll}\hline
1Feature & CE	 &2Feature   &CE	  &3Feature	   & CE	  \\ \hline
aX & 1.5498 & aZ\_aX &0.7176 &aZ\_aX\_endSP & 0.5728\\
spinD &1.6596 & aZ\_spinD &0.9242 &aZ\_aX\_vY0 &0.5848\\
pitN & 1.6598 & aZ\_pitN &0.9828 &aZ\_aX\_startSP &0.5919\\
aZ & 1.8758 &aX\_BL &1.1969 &aZ\_aX\_pitN& 0.6270\\
x0 &2.0851 &aZ\_x0 &1.2286 &aZ\_aX\_z0&0.6684\\
\hline
\end{tabular}%
\caption{Top five feature-sets across three feature settings based on all fastball data of all 12 pitchers.}
\label{ce12pitchers}
\end{table}

\subsection{Fine scale information content of Re-Co dynamics of ${\cal Y}=(pfx_x, pfx_z)$.}
To further look into fine scale information content of ${\cal Y}$, we carry out the MFS-3 and MFS-4 by performing the de-associating calculations with respect to $\{aX, aZ\}$. That is, we subdivide the entire data set with respect to the contingency table $C[aX-vs-aZ]$, as seen in Table~\ref{axaztable}. We perform CE computations for 17 cell-based localities with more than 800 data points. The spin direction ($spinD$) is found as the universal order-1 major factor among the 17 localities. The reason behind this fact can be seen from the homomorphism between the two 3D scatter plots: $(spinD, aX, aZ)$ and $(spinD, pfx_x, pfx_z)$, as seen through their snapshots in Figure~\ref{2Dsnap3Dall}. In Appendix at the link \textrm{https://rpubs.com/CEDA/factorselect}, we clearly see that these two 3D scatter plots are indeed nearly like 2D manifolds, and their homomorphism becomes apparently visible and evident when they are rotated with respect to any 3D directions.

\begin{table}[h!]
\centering
\begin{tabular}{llllll}\hline
$ax$/$az$ & 1 & 2 & 3 & 4&5  \\ \hline
1&675&1612&1233&859&301\\
2&1039&3387&5336&6326&2909\\
3&105&681&2890&2275&835\\
4&943&5451&3906&882&566\\
5&1925&1726&767&215&53\\
\hline
\end{tabular}%
\caption{Contingency table of $C[ax-vs-az]$ based on 12 fastball pitchers' across 2017-2019 seasons. }
\label{axaztable}
\end{table}

\begin{figure}
 \centering
   \includegraphics[width=6in]{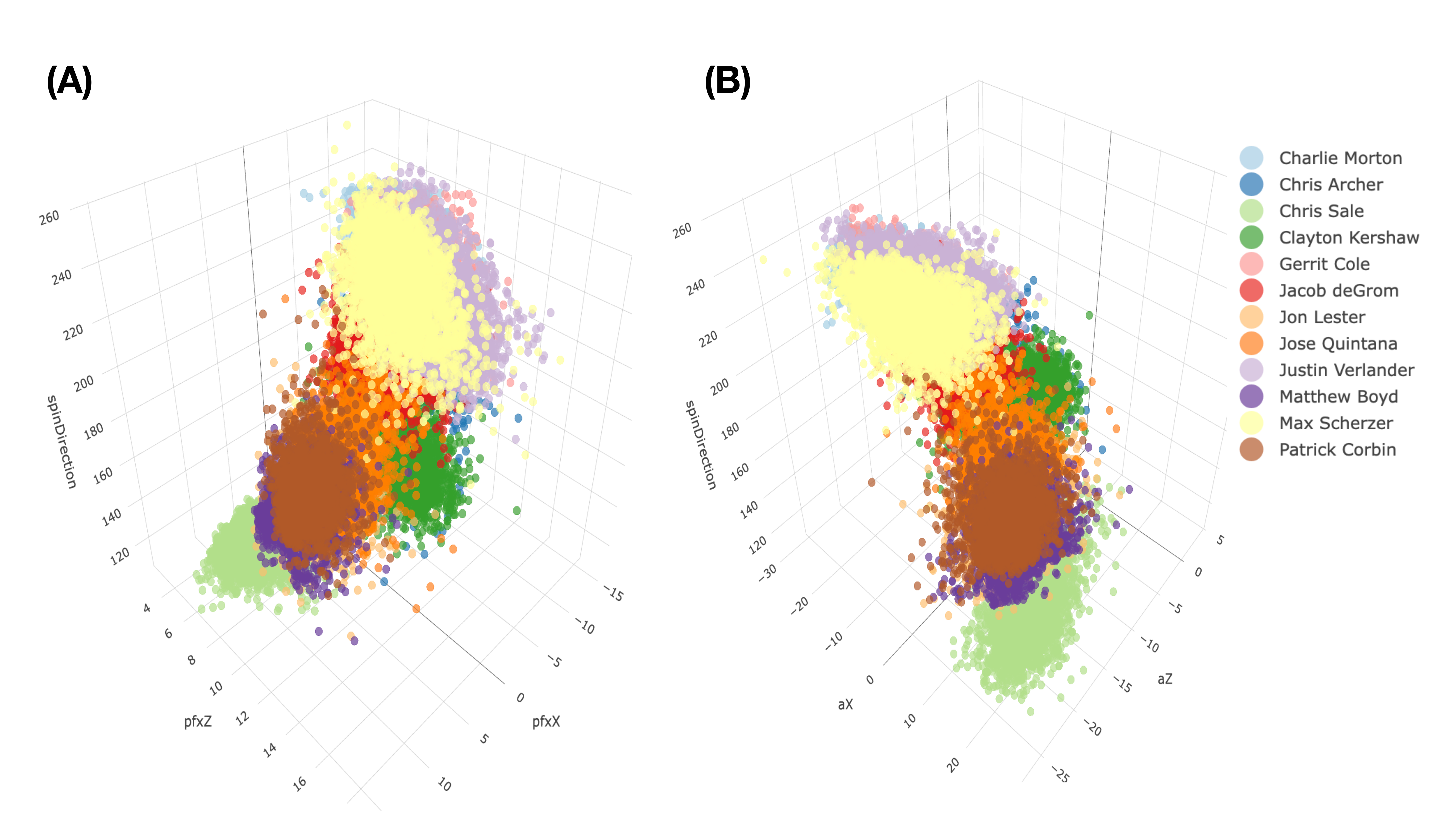}
 \caption{Homomorphic 2D snapshot of 3D manifolds of $(spinD, pfx_x, pfx_z)$ and $(spinD, aX, aZ)$. Also see rotatable 3D manifolds in Appendix (\textrm{https://rpubs.com/CEDA/factorselect}).}
 \label{2Dsnap3Dall}
 \end{figure}

\begin{figure}
 \centering
   \includegraphics[width=6in]{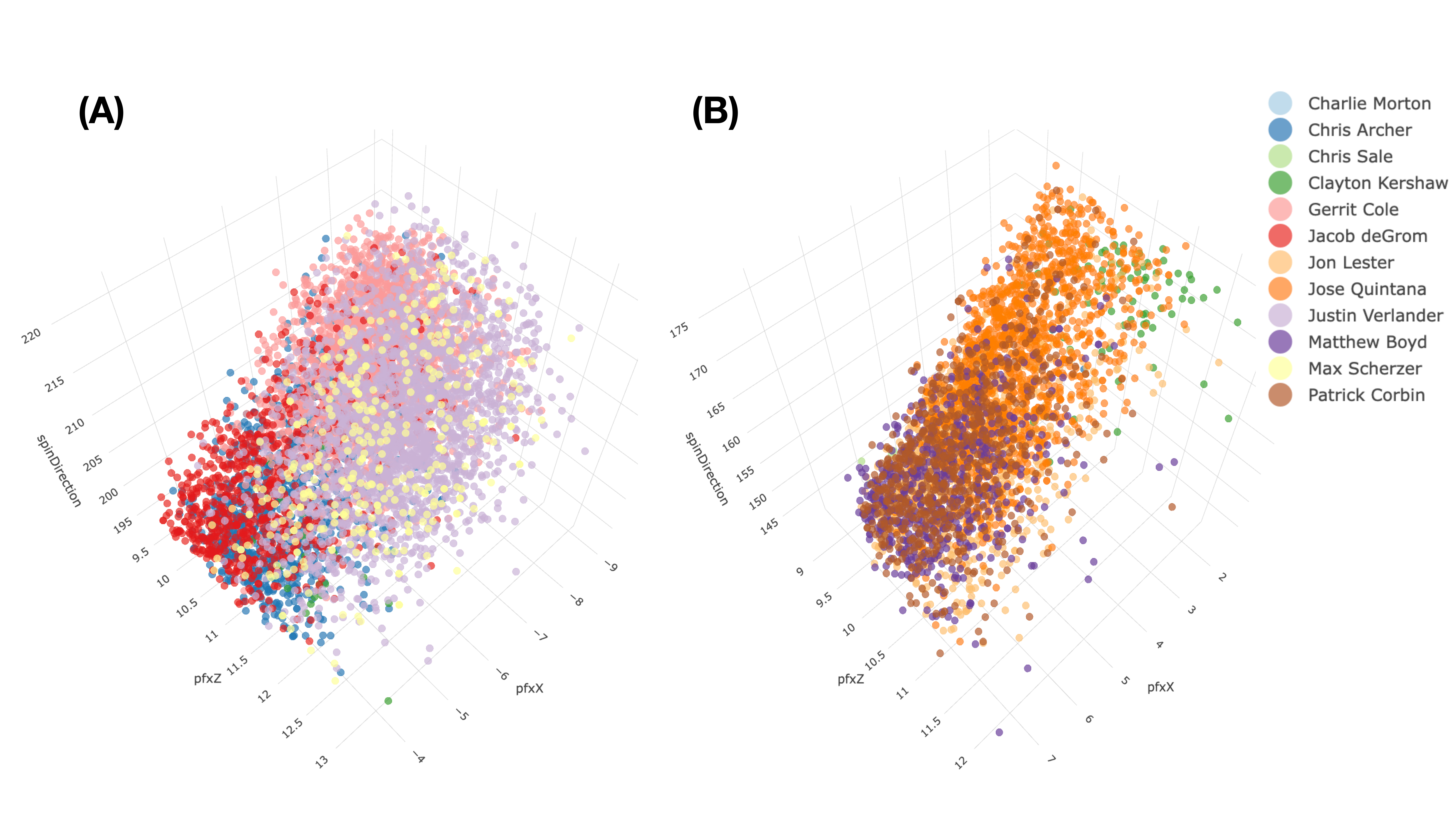}
 \caption{2D snapshot of 3D manifolds of ($spinD, pfx_x, pfx_z$) at the localities: (A)$ (aX, aZ)=(2, 4)$ for right-handed pitchers; (B)$ (aX, aZ)=(4, 3)$ for left-handed pitchers;  Also see rotatable 3D manifolds in Appendix (\textrm{https://rpubs.com/CEDA/factorselect}).}
 \label{2Dsnap3local}
 \end{figure}

Based on this homomorphic relationship between 3D scatter plots of $(spinD, aX, aZ)$ and $(spinD, pfx_x, pfx_z)$, we can visualize these all 25 localities via Figure~\ref{2Dsnap3Dall}, or more precisely through their 3D scatter plots in Appendix. The fact of $spinD$ being an order-1 major factor in all 17 localities can be visualized through two sampled snapshots in Figure~\ref{2Dsnap3local}.

Upon the 17 localities having more than 800 data points, we also report the triplets feature-sets achieving the smallest CE. It is not surprising that all triplets share the common feature-pair $(spinD, endSP)$ with feature $endSP$ standing for ``endspeed''. The most common third member in these triplets is $break-angle (BA)$. These triplets somehow prescribe the fine scale information contents of Re-Co dynamics of ${\cal Y}=(pfx_x, pfx_z)$ upon these localities.

\begin{table}[]
\resizebox{\textwidth}{!}{%
\begin{tabular}{llllll}\hline
$ax$/$az$ & 1 & 2 & 3 & 4&5  \\ \hline
1&NA&BA\_spinD\_endSP&BA\_spinD\_endSP&BA\_spinD\_endSP&NA\\
2&BA\_spinD\_endSP&BA\_spinD\_endSP&BA\_spinD\_endSP&BA\_spinD\_endSP&BA\_spinD\_endSP\\
3&NA& NA&BA\_spinD\_pitN&vX0\_spinD\_endSP&BA\_spinD\_endSP\\
4&BA\_spinD\_endSP&BA\_spinD\_endSP&BA\_spinD\_endSP&vX0\_spinD\_endSP&NA\\
5&BA\_spinD\_endSP&BA\_spinD\_endSP&NA&NA&NA\\
\hline
\end{tabular}%
}
\caption{Table of best triplets within the framework of $C[ax-vs-az]$ based on 12 fastball pitchers' across 2017-2019 seasons. }
\label{axaztable}
\end{table}

In summary, based on Figure~\ref{2Dsnap3local} and 3D plots in Appendix, we can see that the global scale of information content regarding $(aX, aZ)$ coupled with fine scale information content regarding the triplet of features within each locality will facilitate very precise predictive results for ${\cal Y}=(pfx_x, pfx_z)$. At a given locality specified by $(aX, aZ)$, the information of $spinD$ will specify a strip of small region of ${\cal Y}=(pfx_x, pfx_z)$ upon a local piece of manifold of $(pfx_x, pfx_z, spinD)$, as seen in the two panels in Figure~\ref{2Dsnap3local} and their 3D counterparts in Appendix. Further information of two selected features, which together with $spinD$ constitute a triplet, will pinpoint an even smaller region as a precise prediction of ${\cal Y}=(pfx_x, pfx_z)$. This is how inferences of predictive decisions can be made in a complex dynamic system. This data-driven inferential approach based on data's information content is natural and effective.

\subsection{Fine scale information content of local Re-Co dynamics of $Y=pitN$.}
\begin{figure}
 \centering
   \includegraphics[width=6in]{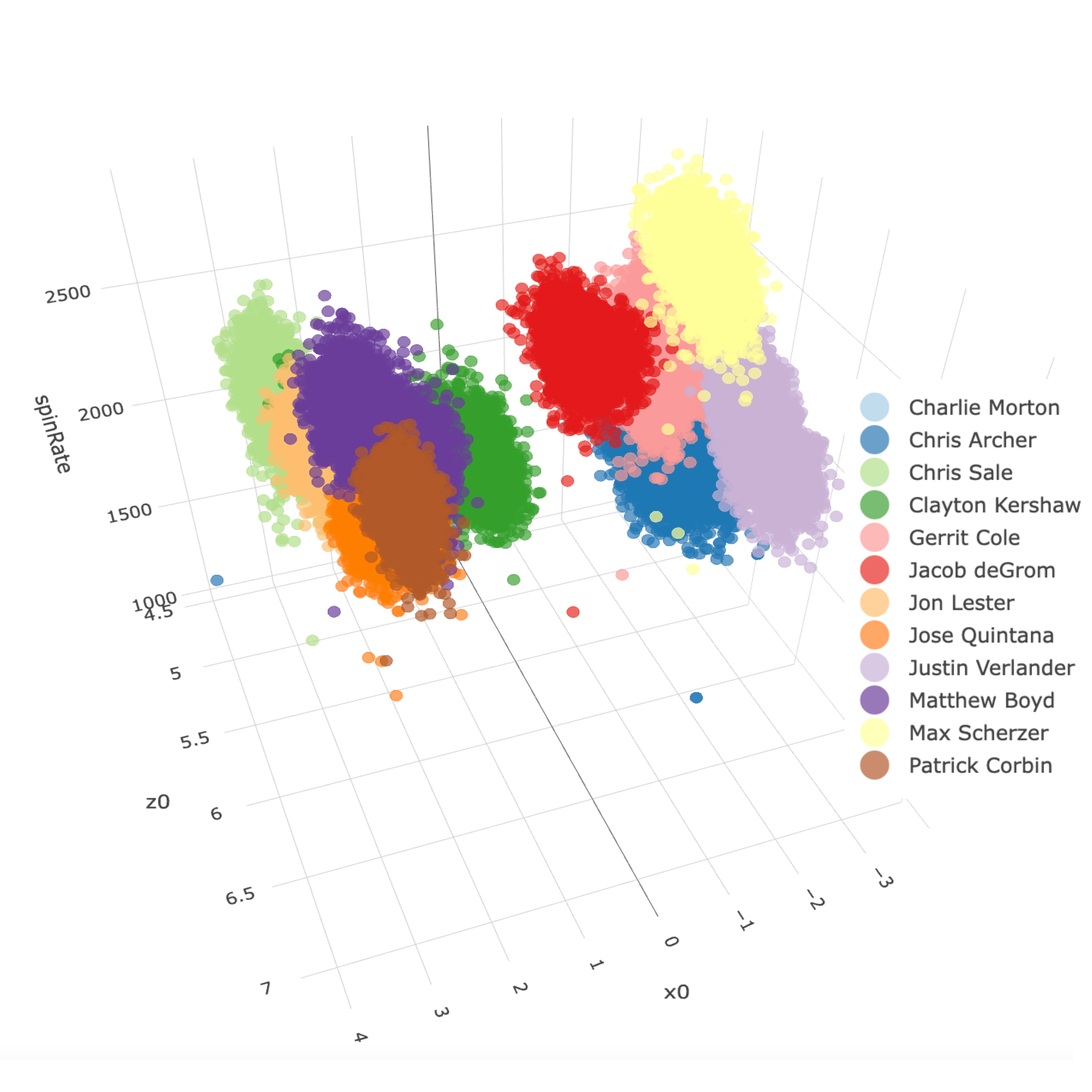}
 \caption{2D snapshot of 3D manifolds of $(x0, z0, spinR)$ based on all the 12 pitchers' pitches data. Also see rotatable 3D manifolds in Appendix (\textrm{https://rpubs.com/CEDA/factorselect}).}
 \label{2Dsnap12whole}
 \end{figure}

In this subsection, we consider the MCC problem that is originally defined with a solo aim of classifying pitches with respect to 12 pitcher-labels based on measurements of 21 features. A usual and direct approach is to build a decision-making base, so-called label-embedding tree based on training data set, see developments and literature review in \cite{FC21} and various approaches, such as Random Forest and variants of Boosting methods, in references there in. Also, another direct approach toward MCC can be performed by taking the pitcher-ID $Y=pitN$ as the response variable and 22 feature as covariate features (including ``season''). We found that the computed best feature-triplet achieving the lowest CE is $(x0, z0, spinR)$. We can see that the effectiveness of $(x0, z0, spinR)$ in solving the MCC on the global scale is indeed not excellent at all, as seen in the 2D projection of the 3D scatter plot of all 12 pitchers in Fig~\ref{2Dsnap12whole}. Indeed, this ineffectiveness of $(x0, z0, spinR)$ is evidently seen through the original 3D plot in Appendix. Though left-handed and right-handed pitchers are separated, the two groups of 6 pitcher-specific (color-coded) point-clouds are overlapping heavily on some regions, while somehow separate in other regions. Therefore, we can see that, if we want to resolve this MCC problem effectively and efficiently, we would need multiscale and heterogeneity perspectives when looking through $(x0, z0, spinR)$.

As a matter of fact, the lowest CE of $(x0, z0, spinR)$ is rather close to the 4 CEs of the rest of top 5 feature-triplets: $(x0, z0, spinD)$, $(z0, spinD, spinR)$, $(x0, z0, aX)$ and $(x0, z0, pfx_x)$. And we know that the 3 features: $pfx_x$, $spinD$ and $aX$, are highly associated across 12 pitchers' pitching dynamics. Since each to-be-classified pitch is a result of a pitcher's pitching dynamics. Wouldn't it be intuitive and reasonable to take a reverse approach by looking at the pitching dynamics first, and then performing the $Y=pitN$ locality specific computations. That is, $Y=pitN$ computations are performed within each locality specified with respect to $(aX, aZ)$. Interestingly, this reverse approach works surprising well.

In this subsection, we present such a natural and effective data-driven MCC resolution developed and constructed from the perspective of complex dynamic system. Based on global Re-Co dynamics of ${\cal Y}=(pfx_x, pfx_z)$, each locality is defined by $(aX, aZ)$. Within each locality, we then consider a local Re-Co dynamics of $Y=pitN$ for the purpose of multiclass classification (MCC) and compute the locality-specific collection of major factors. We report the results in the Table~\ref{axazpitN}.

\begin{table}[]
\resizebox{\textwidth}{!}{%
\begin{tabular}{llllll}\hline
$ax$/$az$ & 1 & 2 & 3 & 4&5  \\ \hline
1&NA&x0\_z0\_spinR&x0\_z0\_spinR&z0\_spinR\_season(*)&NA\\
2&x0\_z0\_season(*)&x0\_z0\_spinR&x0\_z0\_spinR&x0\_z0\_season(*)&x0\_z0\_spinR\\
3&NA& NA&x0\_z0\_spinR(o)&x0\_z0\_spinR(o)&x0\_z0\_spinR\\
4&x0\_z0\_spinR&x0\_z0\_spinR&x0\_z0\_spinR&x0\_z0\_spinR&NA\\
5&aY\_vX0\_x0&x0\_z0\_vX0&NA&NA&NA\\
\hline
\end{tabular}%
}
\caption{Table of best triplets within localities of $(aX,aZ)$ based on 12 fastball pitchers' across 2017-2019 seasons. (*) denoting the (x0, z0, spinR) is ranked next to the best. (o) (*) denoting the (x0, z0, season) is ranked next to the best.}
\label{axazpitN}
\end{table}

\begin{figure}
 \centering
   \includegraphics[width=6in]{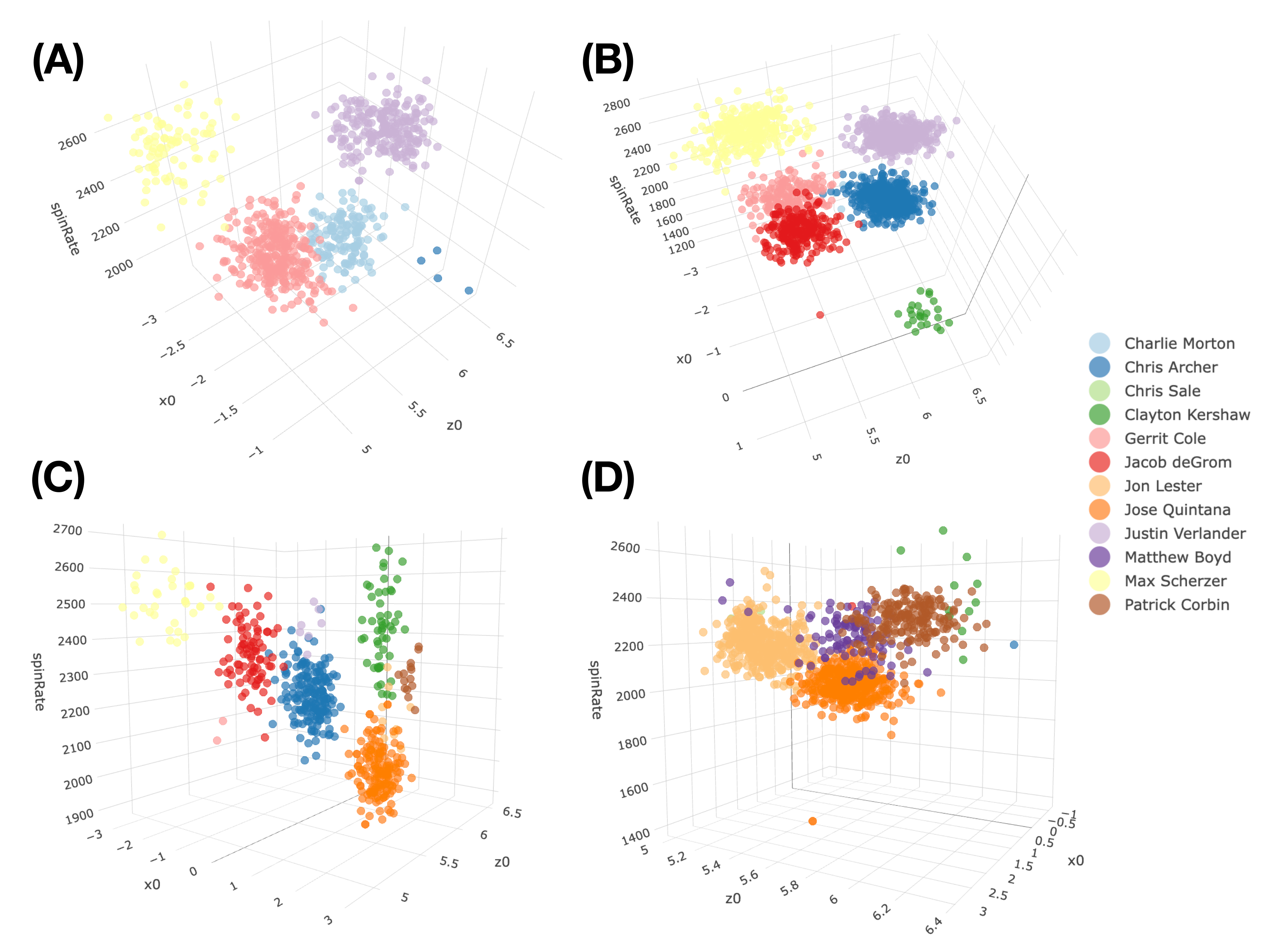}
 \caption{2D project of 3D plot of (x0, z0, spinR) at the localities: (A)$ (aX, aZ)=(1, 2)$; (B) $ (aX, aZ)=(2, 4)$; (C) $ (aX, aZ)=(3, 3)$; (D)$ (aX, aZ)=(4, 3)$.  Also see rotatable 3D manifolds in Appendix  (\textrm{https://rpubs.com/CEDA/factorselect}).}
 \label{2proj3local}
 \end{figure}

Upon each of the 17 localities, a 3D plot is built with respect to the computed feature-triplet. Surprisingly, each feature-triplet provides a nearly perfect separation for all pitchers involved within its corresponding locality. We show 4 panels of  2D projections 4 choices out of 17 (3D) plots in Figure~\ref{2proj3local}. It is noted that the classification results are much more evident through the 3D plots reported in the Appendix.

This surprising resolution of MCC clearly points to the fact that data's authentic multiscale information content should serve as the basis for classification-oriented inferences. Such precise classification inference is interpretable with visible evidences. All evidences commonly converge to an intuitive conjecture: {\bf Build data's information content first and then tackle any complex system related inferential topic issues.} The reason why this conjecture likely hold universally is that most inferential issues are local in nature, which inherently need information of data's multiscale heterogeneity.  Hence, our CEDA-based approach might be the most effective approach to MCC. Our MCC results are further strengthened by the information of three categories of $season$, as would seen in the next subsection.

\subsection{Fine scale information content of Re-Co dynamics of $Y=season$.}
It is imaginable that MLB pitchers' pitching dynamics might change in rather subtle and unknown fashions from season to season. Since such changes are likely very minute and of fine scale. It is likely a difficult, if not an impossible, task to directly discover where these change are located within each individual pitcher's data because the amount of data points is in general limited. We illustrate such difficulties through two of 12 pitchers: Jacob deGrom and Chris Archer.

We first consider Jacob deGrom's case. A direct approach is based on dynamics of $Y=season$ for information about potential changes across 3 seasons in Jacob deGrom's pitching dynamics. We found that the top three triplets in terms of CE are: $(pfx_z, z0, spinD)$, $(pfx_x, vY0, z0)$ and $(vY0, z0, aX)$. These three triplets jointly indicate that the dynamics of ${\cal Y}=(pfx_x, pfx_z)$ is deeply involved. The clear implication of this information is that we need to go into localities defined by $(aX, aZ)$ to search for detailed changes beyond ${\cal Y}=(pfx_x, pfx_z)$ and $(aX, aZ)$. On the other hand, the feature $z0$ seems to be important, so does $vY0$ in dynamics of $Y=season$.

\begin{figure}
 \centering
  \includegraphics[width=6in]{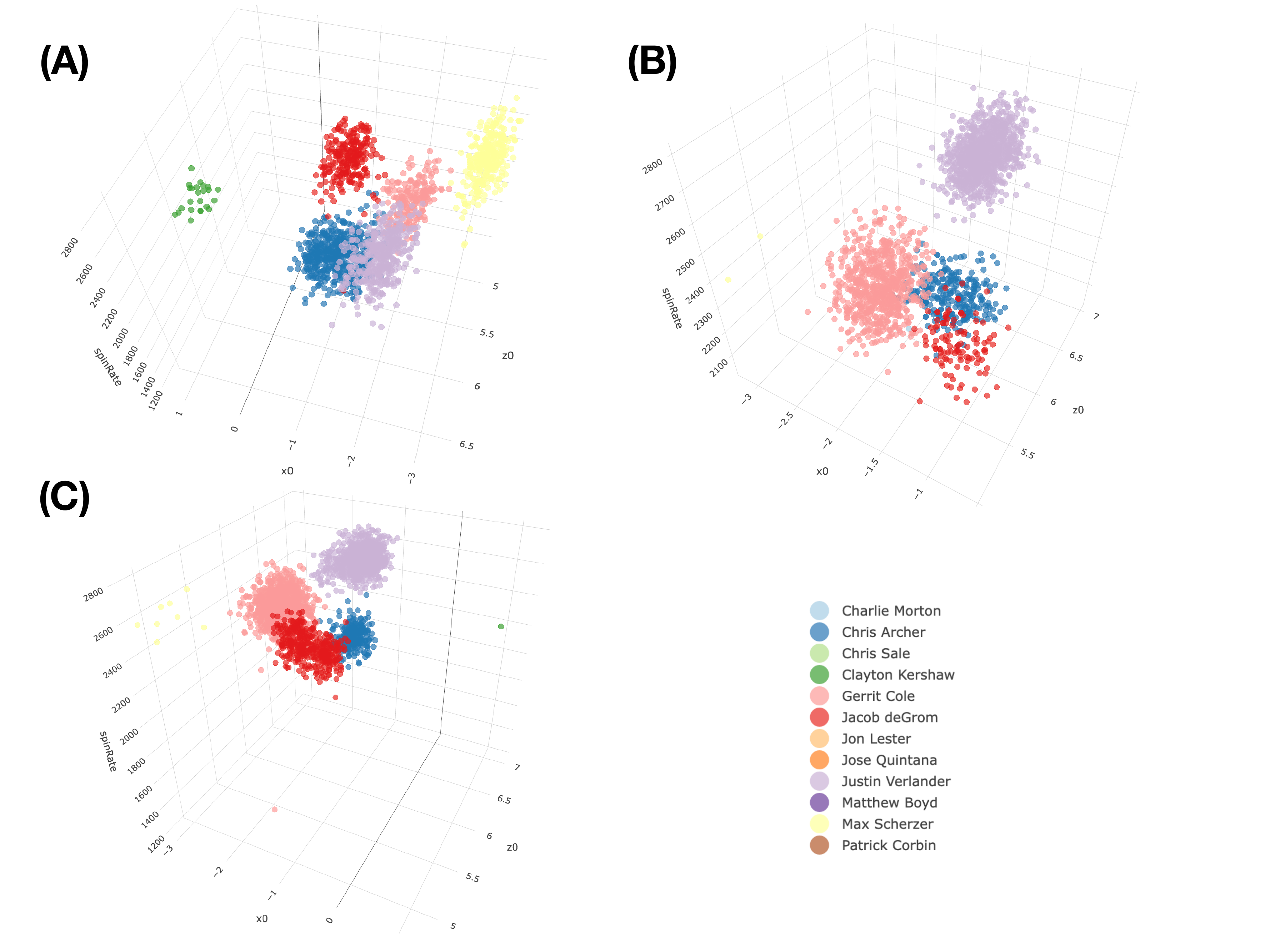}
 \caption{2D project of 3D plot of (x0, z0, spinR) at the locality of $ (aX, aZ)=(2, 4)$: (A) 2017; (B) 2018; (C) 2019.  Also see rotatable 3D manifolds in Appendix  (\textrm{https://rpubs.com/CEDA/factorselect}).}
 \label{2Dseason}
 \end{figure}

As we look through the localities defined by $(aX, aZ)$ with all pitchers, not just the focal one, we indeed can separate the data points with respect to seasons: 2017, 2018 and 2019. We can explicitly see vivid geometries of point clouds of involving pitchers, particularly their relative positions in Euclidean space. It is not entirely expected that, by examining their relative geometric positions across three seasons, the information of changes is already present. For example, as shown through the three panels in Figure~\ref{2Dseason} upon the locality $ (aX, aZ)=(2, 4)$, we witness Max Scherzer's 2017 point cloud (in yellow color) disappeared in 2018 and 2019 almost entirely. Jacob deGrom's 2019 point cloud has significant expansions with respect to the horizontal coordinate of releasing point $x0$. The separations of point clouds belonging to deGrom and Gerrit Cole in 2017 and 2018 are gone. The three Chris Archer's point clouds seem contracting in both $(x0, z0)$ dimensions from 2017 to 2019. In contrast, Justin Verlander's point clouds seem migrate with increasing larger $z0$ values. All these informative patterns are most vividly seen through rotatable 3D plots of the counterpart of Figure~\ref{2Dseason} in Appendix.

Next, we continue on Jacob deGrom as our illustrating focal pitcher in this subsection. Now we go for the direct search based on dynamics of $Y=season$ within locality $ (aX, aZ)=(2, 4)$. We perform our conditional entropy (CE) calculations for three settings: 1-feature to 3-feature. The feature triplet achieves the lowest CE is $(vY0, vZ0, z0)$. We present the 3D plot of Jacob deGrom's point clouds with respect to $(vY0, vZ0, z0)$ across the three seasons. To represent the patterns contained in this 3D plot, we present 4 of its snapshots in four panels of Figure~\ref{2DdeGrom}, see this 3D plot in Appendix. In panel (A), by tuning to a perspective, we see the three season-specific (color-coded) point clouds of Jacob deGrom are evidently located and centered in different locations. In panels (B), (C) and (D), we evidently see their pairwise differences across the three pairs of 3 seasons.  In summary, we can clearly see Jacob deGrom's changes across the three seasons within the locality $ (aX, aZ)=(2, 4)$. Likewise we can explore other localities for similar or distinct changes. Such pieces of information can be very valuable to this pitcher as well as his pitching coach.

\begin{figure}
 \centering
  \includegraphics[width=6in]{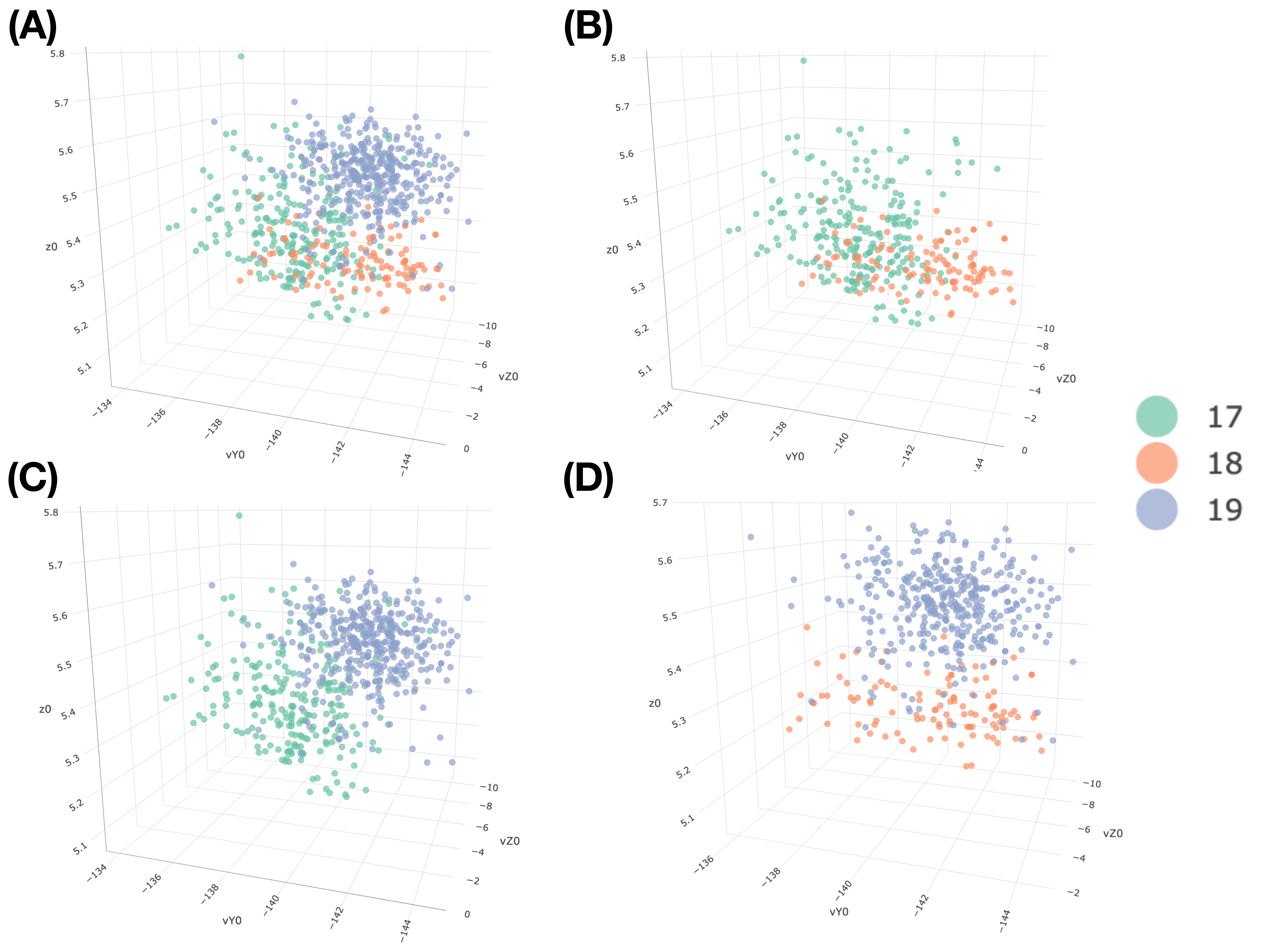}
 \caption{Jacob deGrom's 2D project of 3D plot of $(vY0, vZ0, z0)$ at the locality of $ (aX, aZ)=(2, 4)$: (A) 2017-2019; (B) 2017-2018; (C) 2017-2019; (D)2018-2019.  Also see rotatable 3D manifolds in Appendix  (\textrm{https://rpubs.com/CEDA/factorselect}).}
 \label{2DdeGrom}
 \end{figure}

\begin{figure}
 \centering
 \includegraphics[width=6in]{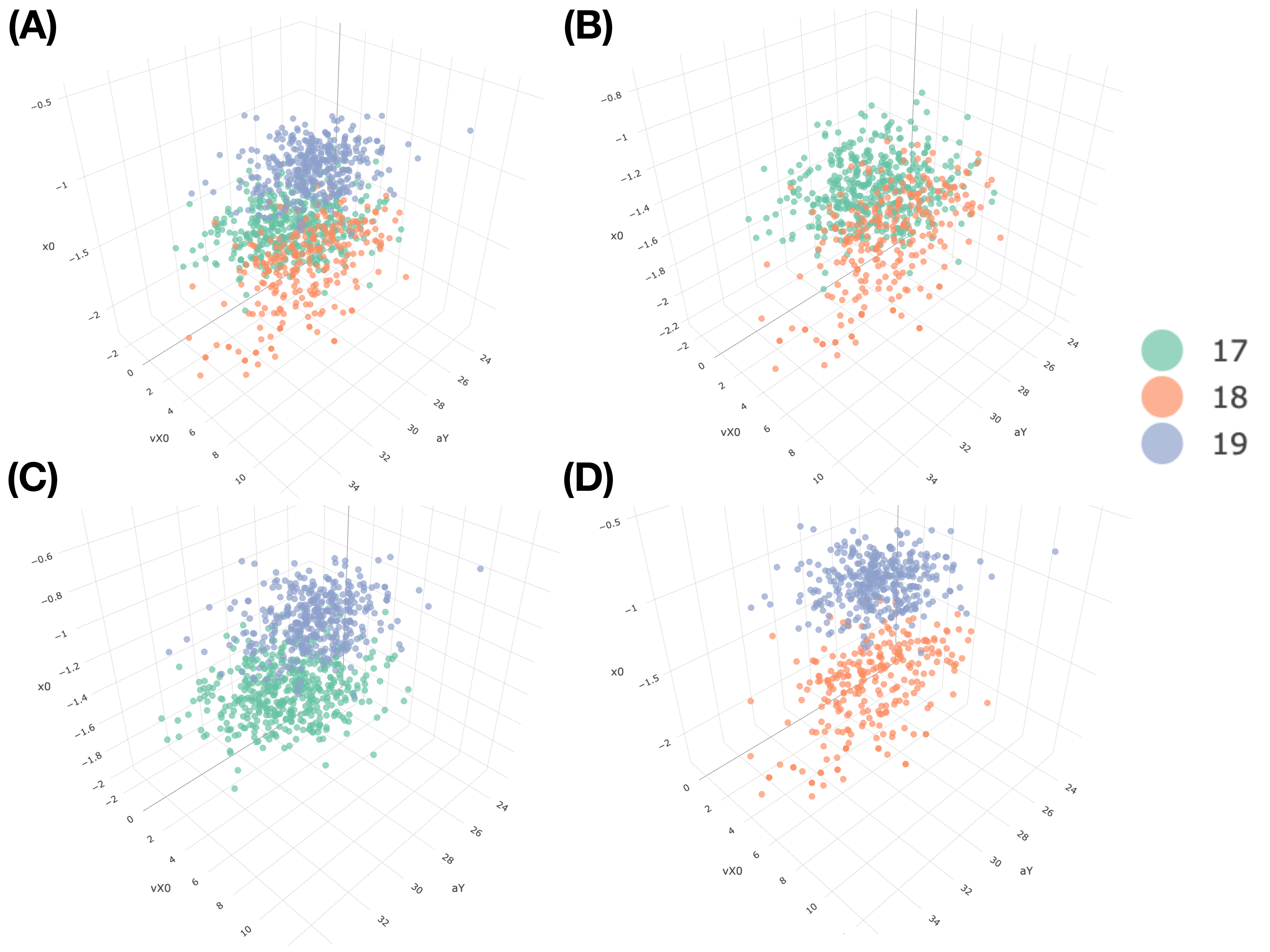}
 \caption{Chris-Archer's 2D project of 3D plot of $(aY, vX0, x0)$ at the locality of $ (aX, aZ)=(2, 3)$: (A) 2017-2019; (B) 2017-2018; (C) 2017-2019; (D)2018-2019.  Also see rotatable 3D manifolds in Appendix  (\textrm{https://rpubs.com/CEDA/factorselect}).}
 \label{2DArcher}
 \end{figure}

In contrast, we look into the locality of $ (aX, aZ)=(2, 3)$ where Chris Archer is an involving pitcher. Again we perform our conditional entropy (CE) calculations with $Y=season$ for three settings: 1-feature to 3-feature. The feature triplet achieves the lowest CE is $(aY, vX0, x0)$, which is very different from the feature-triplet found in Jacob deGrom data. To see the patterns belonging to Chris Archer contained in his 3D plot, which can be found in Appendix, again we present 4 of its snapshots in four panels of Figure~\ref{2DArcher}. In panel (A), by tuning to a perspective, we see the three season-specific (color-coded) point clouds of Chris Archer are evidently located and centered in different locations. In panels (B), (C) and (D), we evidently see their pairwise differences across the three pairs of 3 seasons. That is, we can also clearly see Chris Archer's changes across the three seasons within the locality $ (aX, aZ)=(2, 3)$. In fact, by doing so, we can discover different pitchers' season-specific changes within their pitching dynamics across various localities.

As the final remark of this subsection, it is evident that point clouds belonging to all involving pitchers in localities $ (aX, aZ)=(2, 3)$ and $ (aX, aZ)=(2, 4)$, as shown in Figure~\ref{2DArcher} and Figure~\ref{2DdeGrom} and their counterparts in Appendix, are very well separate from the three perspectives of three seasons. This fact further confirms that MCC topic should be addressed from a complex dynamic system perspective, and its resolution can be a natural by-product of data's multiscale information content with discovered structural heterogeneity.

\section{Conclusions}
From the BRFSS and MLB examples, we vividly see the commonality of structural dependency among features and structural heterogeneity as the collective signatures embedded within large complex systems' structured databases. If they are pertinently extracted and explicitly represented, such signatures would and should lead us to authentic understanding and knowledge embedded within such large systems and further provide resolutions to issues pertaining to the complex systems of interest. In this paper we develop and apply Theoretical Information Measurement based major factor selection (MFS) protocol onto data as one whole as well as onto data in many localities to achieve multiscale information content that contains resolutions to diverse tasks ranging from fundamental inferences to various topic issues. Henceforth, we propose such a computational protocol as a highly adaptable way of studying real-world complex systems.

This data analytic proposal indeed is coherent with our scientific intuition: We need intrinsic knowledge about a targeted complex system in order to derive pertinent resolutions to related topics and problems. The global information derived from the data as one whole serves to define the locality-landscape, which further leads us to traverse through its collection of locality specific pattern information. Such global and local information collectively constitutes the data's information content. That is, our data analytic proposal is sharply contrasting with majority of statistical and machine learning methodologies in literature. If data analysis is for understanding about a targeted complex system and for interpreting all computed results pertaining to this system, then our indirect approach might be more ``natural''.

With thorough illustrations via Example-1 and Example-2, our major factor selection protocol is explicitly reasoned and explained by two operational concepts: ``shadowing'' and ``de-associating'', and their operations are carried out on contingency table platforms. Though these two concepts make explicit and operative use of statistical concept of conditioning of two categorical variables, it is essential to emphasize that resultant conditional variables in general are only conceptually existing, especially in continuous non-Normality cases. They are even not computable with explicit distributional characteristics. However, upon the contingency table platform, we can explicitly and operationally express any results of conditioning with respect to any pair of feature-sets, each of which could be involving multiple 1D variable of any data types. That is, contingency table platform allow us to resolve issues arising from the presence of structural dependency among involving features. Such issues were left unsolved in our previous works on major factor selection in \cite{CCF22a,CCF22b}.

Simultaneously, the ``shadowing'' and ``de-associating'' operational concepts allow us to evaluate and discover what feature-sets offer information beyond any targeted feature-sets at which localities. Given that we can discover major-factor--based localities that contribute exclusive information like no others, our major factor selection protocol demonstrates great potentials and utilities for expand Granger casuality onto any complex dynamic systems with structured data of any types.

At the end, we reiterate a remark about information content of a structured data set or database. What is contained in a structured database? It is beneficial to know the answer to this question in general and conceptual terms. To the authors limited knowledge, the answer highly depends on how annotation is specifically carried out when building such a database. If its curator is one or a group of subject matter scientists, then this structured database is likely encoded with these experts' domain experience and knowledge. The encoding scheme can be simply carried out by picking and choosing ``right'' measurable features. With features' measurements being aligning with domain expertise and understanding, the resultant databases is expected to be encoded with curators' intelligence regarding the complex system of interest. Most importantly, this paper evidently indicates that data analysts could discover knowledge far beyond the curators' encoded intelligence when the complex system of interest indeed embraces with structural dependency and structural heterogeneity.

\vspace{6pt}

\end{document}